\documentclass[fleqn,usenatbib]{mnras}

\usepackage{newtxtext,newtxmath}

\usepackage[T1]{fontenc}

\DeclareRobustCommand{\VAN}[3]{#2}
\let\VANthebibliography\thebibliography
\def\thebibliography{\DeclareRobustCommand{\VAN}[3]{##3}\VANthebibliography}

\usepackage{graphicx}	
\usepackage{amsmath}	

\usepackage{amssymb}	
\usepackage{xcolor}

\newcommand{\au}{\,\mathrm{au}}
\newcommand{\yr}{\,\mathrm{yr}}
\renewcommand{\deg}{^\circ}
\newcommand{\hmag}{\,\mathrm{mag}}

\title[Minimum $q$ and dwell times of NEAs]{Minimum perihelion distances and associated dwell times for near-Earth asteroids}

\author[A. Toliou et al.]{
Athanasia Toliou,$^{1}$\thanks{Corresponding author; email: athanasia.toliou@ltu.se}
Mikael Granvik,$^{1,2}$
Georgios Tsirvoulis$^{1}$
\\
$^{1}$Asteroid Engineering Laboratory, Space Systems, Lule\r{a} University of Technology, Box 848, SE-98128 Kiruna, Sweden\\
$^{2}$Department of Physics, PO Box 64, 00014 University of Helsinki, Finland
}

\date{Accepted 2021 July 2. Received 2021 June 29; in original form 2021 May 20}

\pubyear{2021}

\begin{document}
\label{firstpage}
\pagerange{\pageref{firstpage}--\pageref{lastpage}}
\maketitle

\begin{abstract}
The observed near-Earth asteroid population contains very few objects with small perihelion distances, say, $q\lesssim0.2\au$. NEAs that currently have orbits with larger $q$ might be hiding a past evolution during which they have approached closer to the Sun. We present a probabilistic assessment of the minimum $q$ that an asteroid has reached during its orbital history. At the same time, we offer an estimate of the dwell time, that is, the time $q$ has been in a specific range. We have re-analyzed orbital integrations of test asteroids from the moment they enter the near-Earth region until they either collide with a major body or are thrown out from the inner Solar System. We considered a total disruption of asteroids at certain $q$ as a function of absolute magnitude ($H$). We calculated the probability that an asteroid with given orbital elements and $H$ has reached a $q$ smaller than a given threshold value and its respective dwell time in that range. We have constructed a look-up table that can be used to study the past orbital and thermal evolution of asteroids as well as meteorite falls and their possible parent bodies. An application to 25 meteorite falls shows that carbonaceous chondrites typically have short dwell times at small $q$, whereas for ordinary chondrites it ranges from $10\,000$ to $500\,000$ years. A dearth of meteorite falls with long dwell times and small minimum $q$ supports a super-catastrophic disruption of asteroids at small $q$.
\end{abstract}



\section{Introduction}

The orbital history of near-Earth objects (NEOs) is often described as chaotic, since it is characterized by close encounters with the terrestrial planets, as well as various resonant phenomena. Although the number of observed  NEOs that exhibit orbits with small perihelion distances ($q$) is limited, some members of the NEO population may have had a very different past and may have acquired very small $q$ values. One definition of small $q$ can be given by considering the $q$ at which the theoretically-predicted number of NEOs and the observed NEOs start diverging significantly \citep{Granvik2016}. Based on that definition, we denote as NEOs with small $q$, those with $q<0.2\au$.

The majority of asteroids orbiting in the near-Earth region end their evolution by falling into the Sun \citep{Farinella1994,Gladman2000,Marchi2009}. Their demise can be expedited as NEOs that reach below some critical perihelion distance (as a function of their size) can be completely destroyed \citep{Granvik2016}. However, this evolutionary trajectory towards the Sun is not monotonous; having a way to assess the minimum perihelion distance that an asteroid has reached can be helpful in identifying which of the observed asteroids have had at some point in the past an orbit that came close to the Sun. Consequently, these objects can be studied further in order to determine their dynamical and physical properties (such as shape, spin rate, spectrum, albedo, etc.) and study their thermal histories. 

Meteoritic samples show signs of heating processes \citep{Keil2000}, which are typically attributed to the decay of short-lived radionuclides \citep{Grimm1993} and shock heating by collisions \citep{Jutzi2020}. However, to some degree, the heating can be a result of extreme solar irradiation \citep{Marchi2009}. Since the heliocentric orbit of a meteorite fall, or a fireball in general, is expected to be similar to the orbit of its immediate parent body, knowing the early orbital evolution of the latter is crucial in order to determine several physical properties of the former, such as the maximum temperature it has experienced, which can in turn be useful in estimating its composition and compare the results with mineralogical studies in the laboratory.

\citet{Marchi2009} explored this possibility by studying the dynamical history of NEOs and deriving their surface temperatures by using thermophysical models. In their study, they used the NEO population model by \citet{Bottke2000,Bottke2002} which was the most complete model at that time. Since then, an updated model \citep{Granvik2016,Granvik2017,Granvik2018} has been derived, that makes some improvements on the \citet{Bottke2000,Bottke2002} model. Most importantly, the new model accounts for the disruption of NEOs close to the Sun.

The goal of the present work is to revisit the study of the dynamical past of the asteroidal component of the NEOs, the near-Earth asteroids (NEAs), and construct a look-up table which can offer a probabilistic assessment of the history of the evolution of $q$ for any asteroid with given orbital elements $a,e,i$ and absolute magnitude $H$. Using the same data set as \citet{Granvik2017,Granvik2018} we will be able to provide the probability the asteroid has at some point reached a perihelion distance below a given threshold value $(q_s)$ and, in addition, provide information concerning the cumulative time that it has spent having $q$ in a specific range.

\section{Definitions and Methods}\label{sec:methods}

\subsection{Definitions and simulation data}

\begin{figure}
\centering
\includegraphics[width=0.48\textwidth]{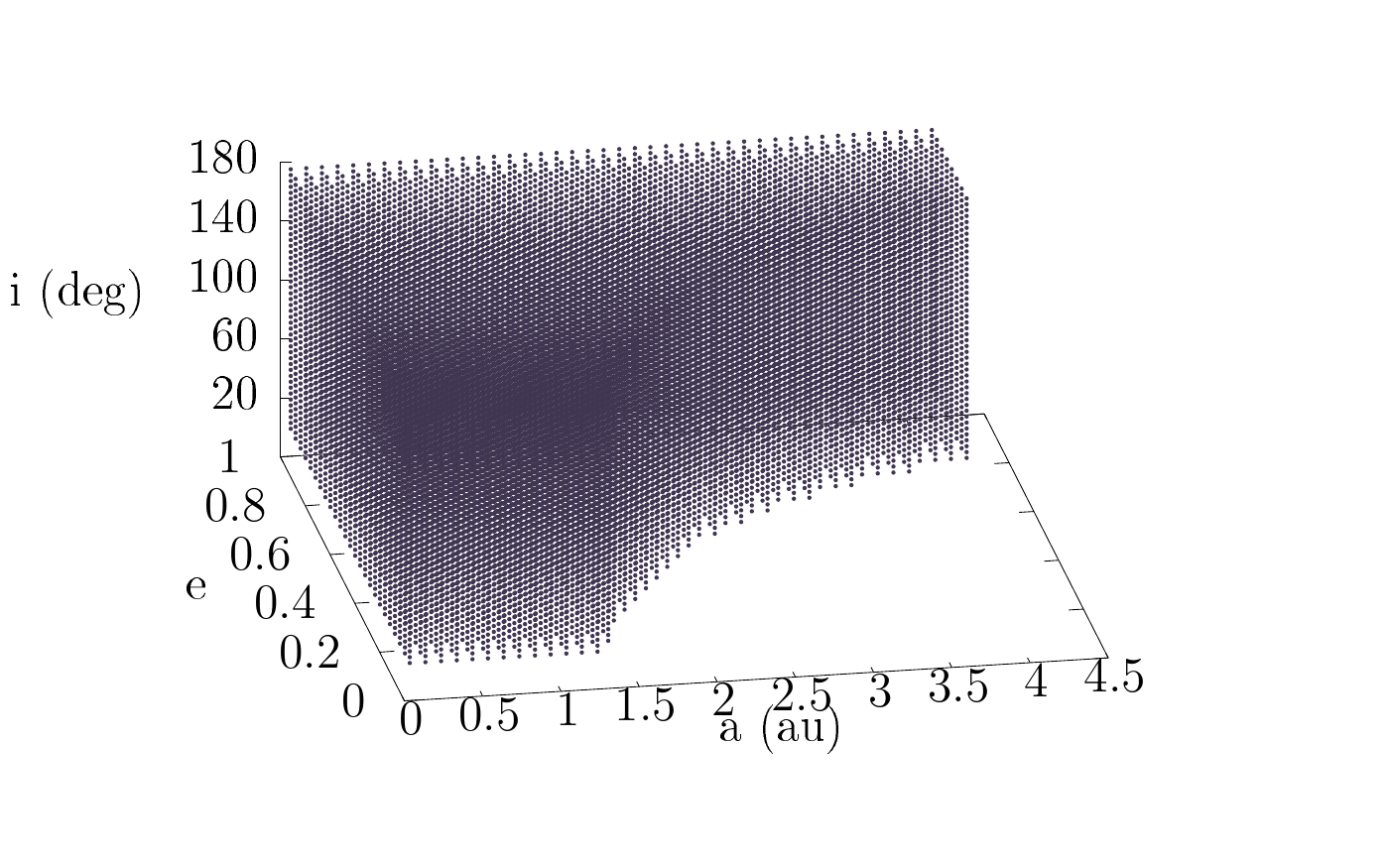}
\caption{The volume in orbital elements space populated by NEAs, split in 42 bins in $a$, 25 bins in $e$ and 45 bins in $i$. To each point shown in the grid correspond 40 bins in H, leading to a total resolution of 1,890,000 cells from which we exclude all cells with $q>1.3\au$. For every cell we calculate the probability that an asteroid with corresponding $a$, $e$, $i$ and $H$ has had at some point in its past evolution $q$ lower than each of the 26 threshold values $q_s$ and the associated dwell times.}
\label{fig:grid}
\end{figure}

Following the \citet{Granvik2018} notation, we define as NEA an object having $q<1.3\au$ and $a<4.2\au$. We divide the near-Earth region in 42 semimajor-axis bins of width $0.1\au$ in the range $0<a<4.2\au$, 25 eccentricity bins of width $0.04$ in the range $0<e<1$, 45 inclination bins of width $4\deg$ in the range $0\deg<i<180\deg$ (to include both prograde and retrograde asteroids), and 40 absolute-magnitude bins of width $0.25$ in the range $15<H<25\hmag$, resulting in a grid with $1,890,000$ cells. From this grid we excluded the cells for which $q>1.3\au$ \autoref{fig:grid}. Next, we split the perihelion distances $0<q<1.3\au$ in 26 bins of width $0.05\au$ with the upper bin boundaries described by $q_s=0.05,0.1,0.15,\ldots,1.3\au$. We use the same bin widths and parameter ranges as \citet{Granvik2018} in order to match the public low-resolution version of their model.

The input data for our work contain the full integrations of $\sim70.000$ test asteroids, performed by \citet{Granvik2018}, from the moment they entered the near-Earth region from the main asteroid belt (MAB), until they ended up in their respective sinks, that is, the location reached by a given test asteroid when the
integration was stopped, which was forced if the heliocentric distance of the asteroid became smaller than the solar radius $R_{\odot}=4.65 \times 10^{-3}\au$ or its $a>100\au$. The time interval at which the orbital elements of a test asteroid were recorded (the output timestep) is 250~yr. To assess if the time resolution is adequate for our needs, we took a sample of 100 random test asteroids and computed the cumulative distribution of the change in $q$ between consecutive timesteps. We found that 95 per cent  of the differences are $\lesssim0.05\au$. We note that \citet{Granvik2018} did a similar analysis in the ($a,e,i$) space.

\subsection{Accounting for super-catastrophic disruption}

The orbital integrations of test asteroids in the near-Earth region, that we utilize here and that were utilized for the construction of the NEO population model \citep{Granvik2018}, do not account for size-dependent effects such as Yar\-kov\-sky and the super-catastrophic disruption at small $q$. The former was shown to be negligible compared to the effects of close planetary encounters and the latter was taken into account as a post-processing step.

Here we want to explicitly take into account the size dependence of super-catastrophic disruption at the level of the orbital evolution of individual test asteroids, because it affects the statistics of prior orbital evolution. We consider this disruption mechanism a physical limitation and the only constraint that prevents NEAs from having orbits with very small $q$.

To achieve this, we assigned all 40 $H$ values to each test asteroid and treated each $H$ scenario individually. For the correlation between $H$ and the average disruption distance ($\bar{q}_*$), \citet{Granvik2016} provide us with an estimate of $\bar{q}_*$ for three different $H$ ranges: $17<H<19\hmag$, $20<H<22\hmag$, and $23<H<25\hmag$. Assuming a similar trend in $\bar{q}_*$ in the range $15<H<17\hmag$, we can determine $\bar{q}_*$ for each one of the 40 $H$ bins in our grid. We approximate $\bar{q}_*$ as a function of $H$ with $\bar{q}_*(H)=0.02H - 0.3$, which is an accurate enough approximation considering the uncertainties associated with the estimation of $\bar{q}_*$. Finally, to account for the disruption of asteroids with small $q$, we ignore the test asteroids' subsequent evolution after the first time their $q\leq q_*(H)$. Note that for the first $H$ bin ($15.0<H<15.25\hmag$) this constraint is irrelevant because the disruption distance would be smaller than the radius of the sun, therefore the integration would have already been terminated. 

\subsection{Minimum perihelion distance reached by an asteroid}\label{sub:qin}

Let us then look at how we compute the probability that a given NEA with orbital elements and H within the range of a cell has had $q<q_s$ during its orbital history.
Our algorithm works on an object by object and timestep by timestep basis. For one timestep in one asteroid's orbital history, we locate the $(a,e,i,H)$ cell it belongs to, and increase the event counter corresponding to the $(a,e,i,H)$ cell by one. 

The next step is to find the minimum $q$ value in the interval from the beginning of the test asteroid's orbital integration, until that particular timestep. We then add one to all $q$ counters $(a,e,i,H,q_s)$ that have $q_s$ equal to or larger than the recorded minimum $q$ value.

We follow the same procedure for all available test asteroids and for all output timesteps. Next, we add up all the $q$ counters and event counters for individual test asteroids separately for each ($a,e,i,H$) cell. By dividing the total sum in each $q$ counter with the total sum in the corresponding event counter, we get the probability that an object with given orbital elements has had a $q<q_s$ at some point in its past:

\begin{equation}
    p_{q_s}\left(a,e,i,H\right)=\frac{N_{q_s}\left(a,e,i,H\right)}{N\left(a,e,i,H\right)},
\end{equation}
where $N_{q_s}$ is the total sum in each $q$ counter and $N$ the total sum in the event counters for every $(a,e,i,H)$ cell.

\subsection{Accounting for contributions from different source regions or escape routes}

The integrations used above are not properly weighted. For example, the initial conditions for the integrations were obtained by using different size ranges in the inner and the outer MAB \citep{Granvik2017}. Hence, to reduce biases caused by the choice of initial conditions, we need to take into account the different contribution from every escape region (ER) in the MAB to the NEA population \citep{Granvik2018}. To this end, we calculate the ER-specific $p_{q_{s\text{ER}}}(a,e,i,H)$ to end up with the final distribution of probabilities, weighted by the contribution of each ER.

For the computations, we split the total number of test asteroids in six groups, according to their recorded escape routes from the MAB as defined by \citet{Granvik2017}. In particular, we use i) 8128 asteroids from the Hungaria and ii) 8309 asteroids from the Phocaea families, iii) 11545 asteroids that have escaped through the 3:1 MMR with Jupiter that also incorporated the outer part of the $\nu_6$ secular resonance, iv) 11983 from the 5:2 MMR complex including not only 5:2 but also the 8:3 and 7:3 MMRs, v) 6826 from the 2:1 MMR complex, that considers the contribution of the 11:5 and 9:4 MMRs as well as the z2 secular resonance and finally, vi) 19701 asteroids that escaped throught the $\nu_6$ secular resonance complex, which includes also the 4:1 and 7:2 MMRs.

The \citet{Granvik2018} NEO population model provides us with the relative fraction of NEOs from each ER that contribute to each cell, $\beta_{\text{ER}}(a,e,i,H)$. Since we only take into account asteroids that entered the near-Earth region though escape routes from the MAB, we  exclude the contribution from Jupiter-family comets (JFC) and re-normalize $\beta_{\text{ER}}$.

The linear combination
\begin{equation}
    P_{q_s}\left(a,e,i,H\right)=\sum_{\text{ER}=1}^6 \beta_{\text{ER}}\left(a,e,i,H\right) p_{{q_s}_\text{ER}}(a,e,i,H)
\end{equation}
gives us the final weighted probabilities in each $(a,e,i,H)$ cell. 

An important point to be raised is that the ER probabilities  $\beta_{\text{ER}}$ are derived by taking into account the entire population of test asteroids. Each time a test asteroid crosses the average disruption distance corresponding to its size, there will be a mismatch between the source specific probabilities and the contribution of this ER to the NEA population. Consequently, the linear combination of some $p_{{q_s}_\text{ER}}(a,e,i,H)$ multiplied with the respective $\beta_{\text{ER}}(a,e,i,H)$ might lead to inexact results. Especially for cells that are not very populated, the weighted probabilities do not add up to unity for the outermost $q$ counter. To overcome this problem, we manually set $\beta_\text{ER}=0 $ for every cell in which a discrepancy like the one described above occurs, and re-normalized the $\beta$ for the rest of the ERs. 

\subsection{Dwell times}\label{sub:dwell}

A similar approach is used to derive the dwell times in each $q$ bin defined by the $q_s$ values, i.e., the time an asteroid has $q$ in that range. At one timestep and for one test asteroid, we locate the ($a,e,i,H$) cell it belongs to and record the number of times its $q$ falls into any $q$ counters, as described by the $q_s$ values. This takes into account the orbital evolution from the beginning of the integration of this object until that point in time. If during the orbital evolution of the particle an ($a,e,i,H$) cell is visited multiple times, all previous recordings for that cell are erased and substituted by the last occurrence to avoid overlapping counts. The actual dwell time is found by multiplying the counts in each counter by 250~yr.

After dividing all test asteroids into groups according to their re\-spe\-ctive ERs, we calculate the ER-specific average $\bar{\tau}_{{q_b}_\text{ER}}(a,e,i,H)$ and median $\tilde{\tau}_{{q_b}_\text{ER}}(a,e,i,H)$ dwell times, re\-cor\-ded in every ($a,e,i,H$) cell. Their linear combination after multiplying with the re-normalized $\beta_{\text{ER}}$ gives us the weighted, average, and median dwell times. 

\begin{equation}
\begin{split}
    \bar{T}_{q_b}\left(a,e,i,H\right)&=\sum_{\text{ER}=1}^6 \beta_{\text{ER}}\left(a,e,i,H\right) \bar{\tau}_{{q_b}_\text{ER}}(a,e,i,H)
    \\
    \tilde{T}_{q_b}\left(a,e,i,H\right)&=\sum_{\text{ER}=1}^6 \beta_{\text{ER}}\left(a,e,i,H\right) \tilde{\tau}_{{q_b}_\text{ER}}(a,e,i,H)
\end{split}
\end{equation}

\subsection{Uncertainties}\label{subsec:uncertainties}

In order to understand the statistical uncertainties involved in our calculations, we divide the test asteroids in two groups, according to their integration designations -- even and odd numbers --, which should give a random enough division.

We repeat a similar process as described in section~\ref{sec:methods} for both groups. After calculating $P_{q_s}\left(a,e,i,H\right)$, $\bar{T}_{q_b}\left(a,e,i,H\right)$ and  $\tilde{T}_{q_b}\left(a,e,i,H\right)$, we subtract the values of the even-numbered group from the odd-numbered group. 

For the uncertainty in $P_{q_s}\left(a,e,i,H\right)$ this approach is sufficient. However, in the case of the dwell times, even a proportionally small difference between the even and odd population values can be a large number, if the even and odd population values are already large. Therefore, we normalize the absolute value of the difference by dividing with the nominal dwell times coming form the entire sample of test asteroids. The resulting number, which can be larger that unity, allows us to get a sense of the degree of uncertainty of the calculated dwell times.

\section{Results}

\subsection{Minimum perihelion distance}

In what follows, we will present a few representative cases, make comparisons between the distributions in various cells, and verify that our results are in accordance with what is expected. This will also serve as a validation of our methodology.

\begin{figure*}
\centering
\includegraphics[width=0.49\textwidth]{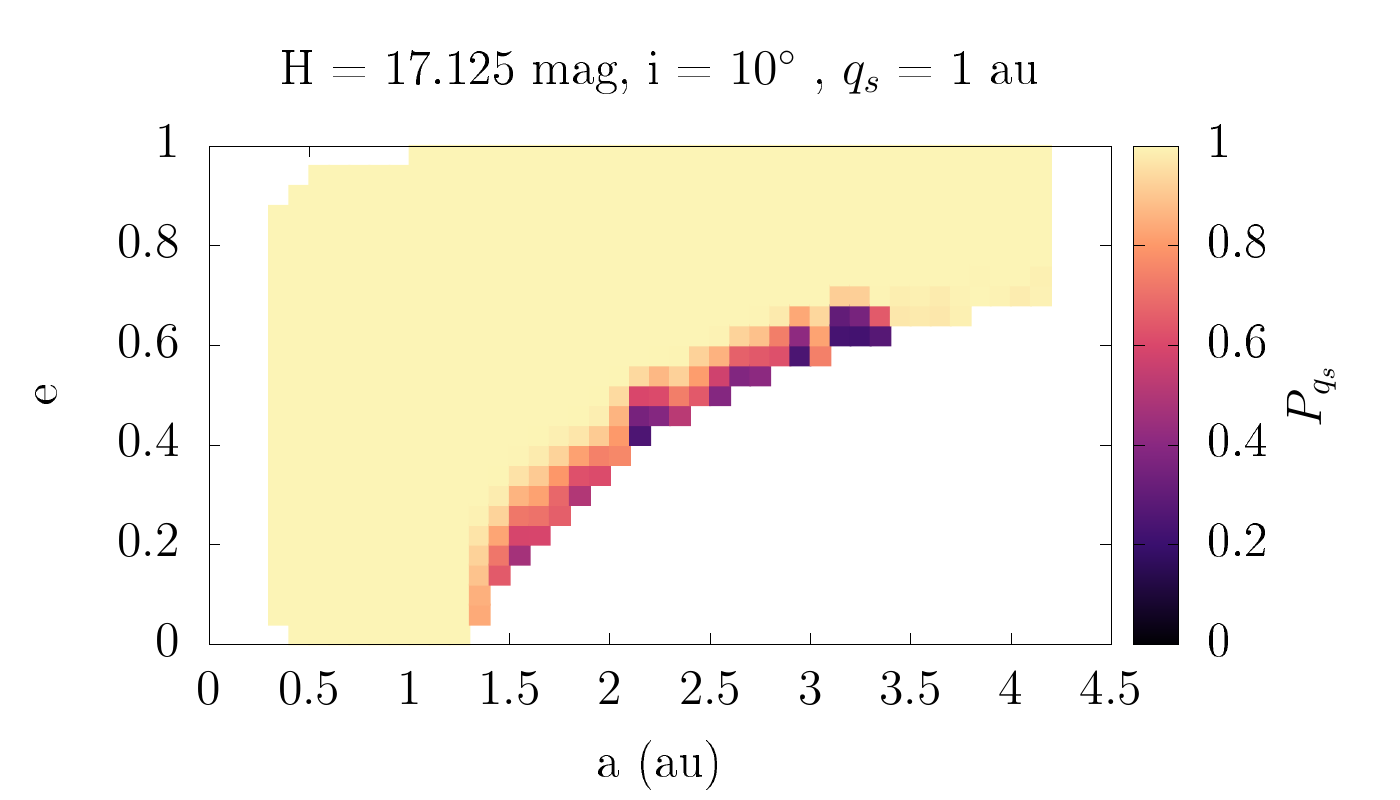}
\includegraphics[width=0.49\textwidth]{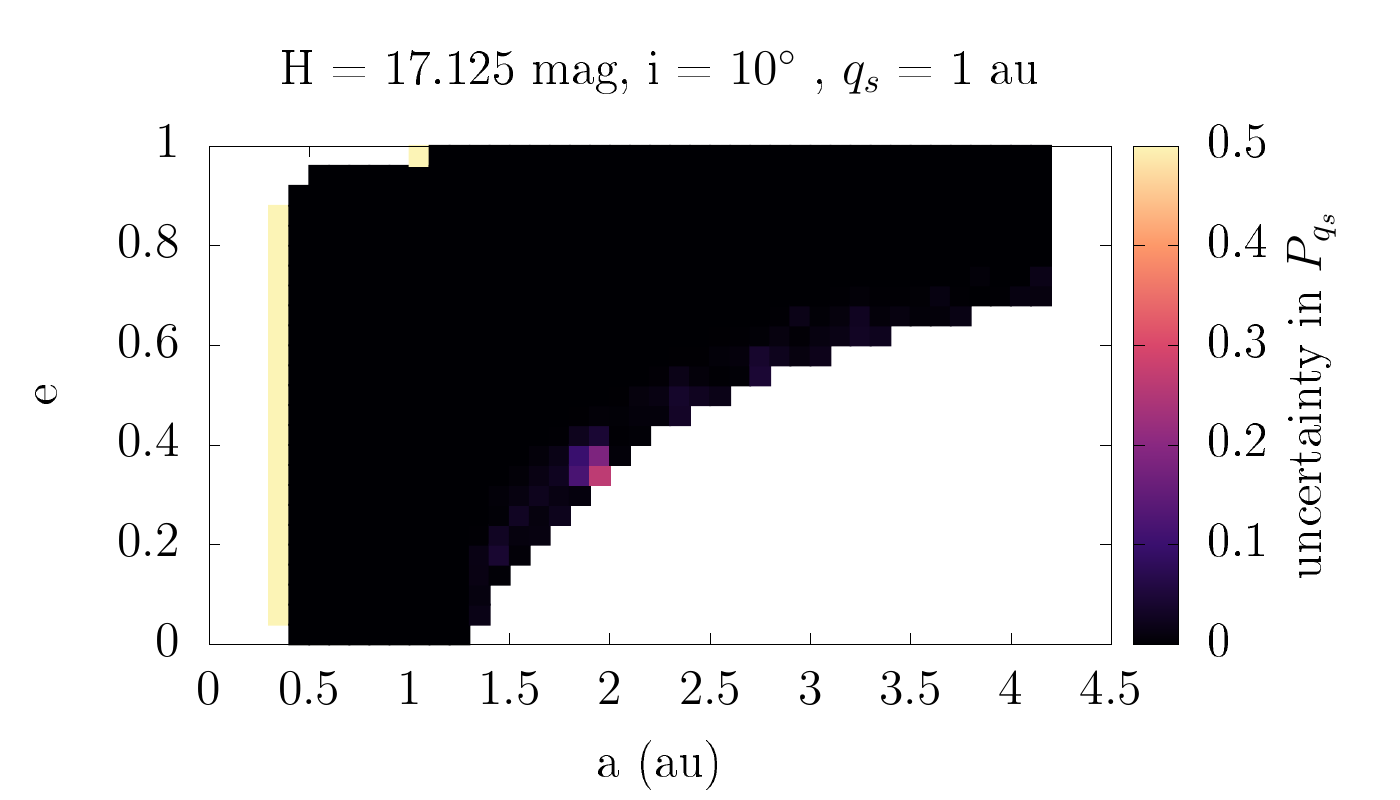}

\includegraphics[width=0.49\textwidth]{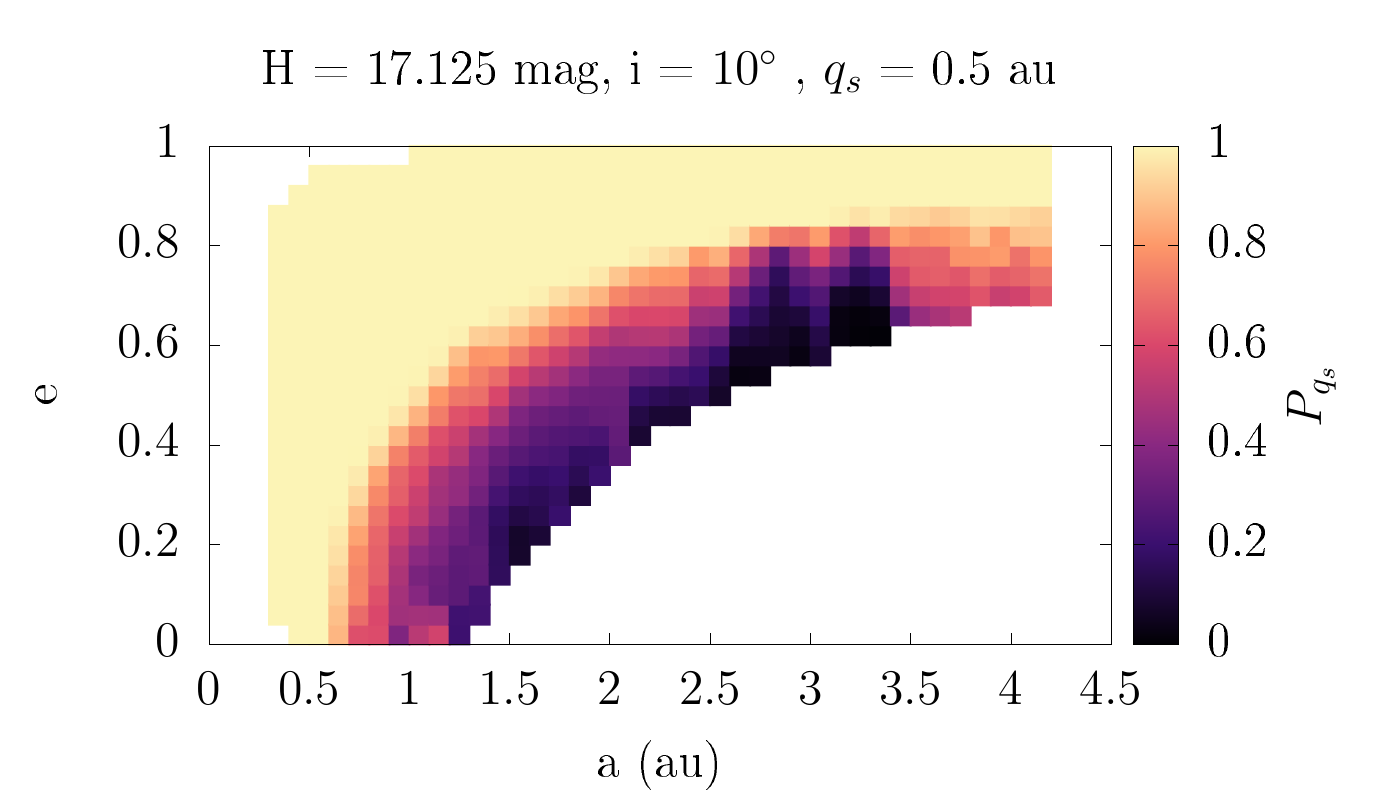}
\includegraphics[width=0.49\textwidth]{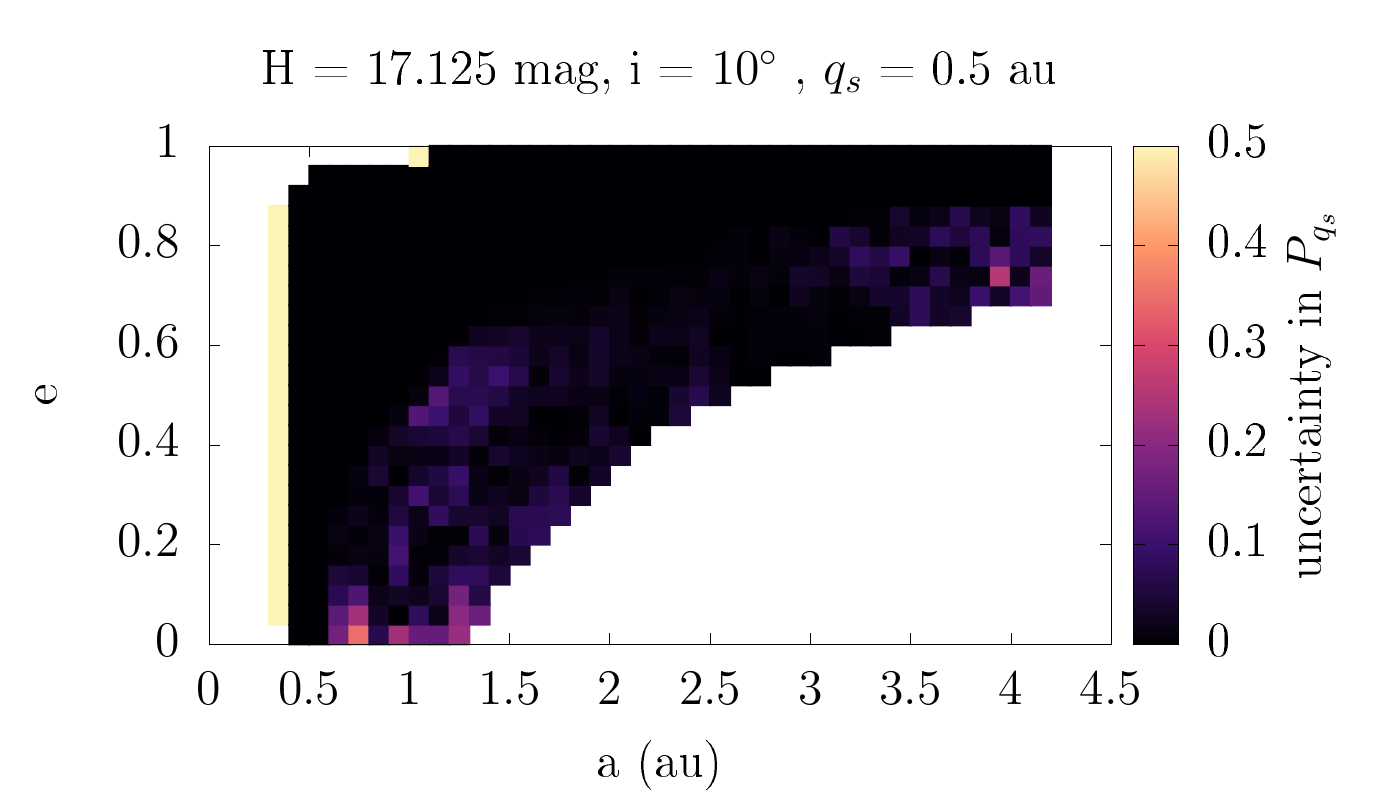}

\includegraphics[width=0.49\textwidth]{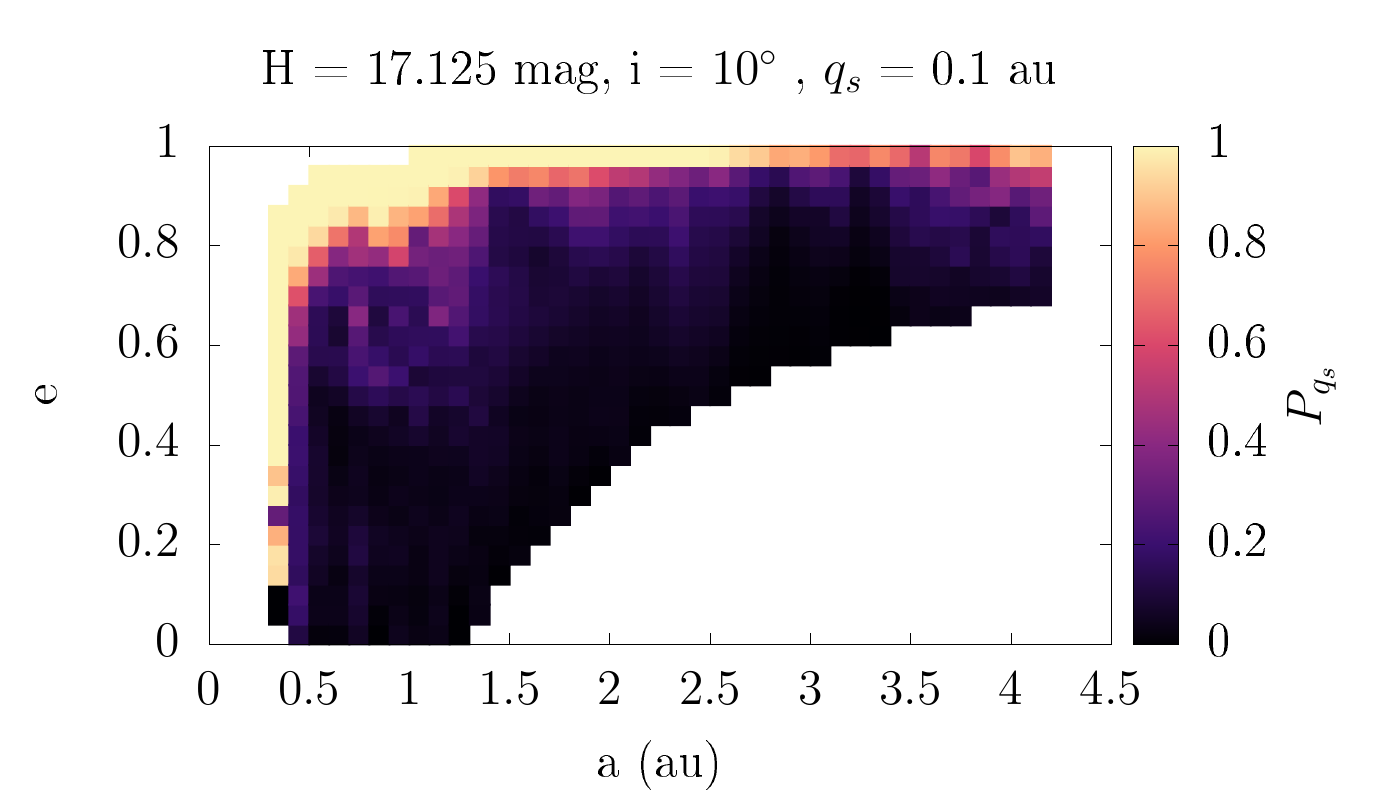}
\includegraphics[width=0.49\textwidth]{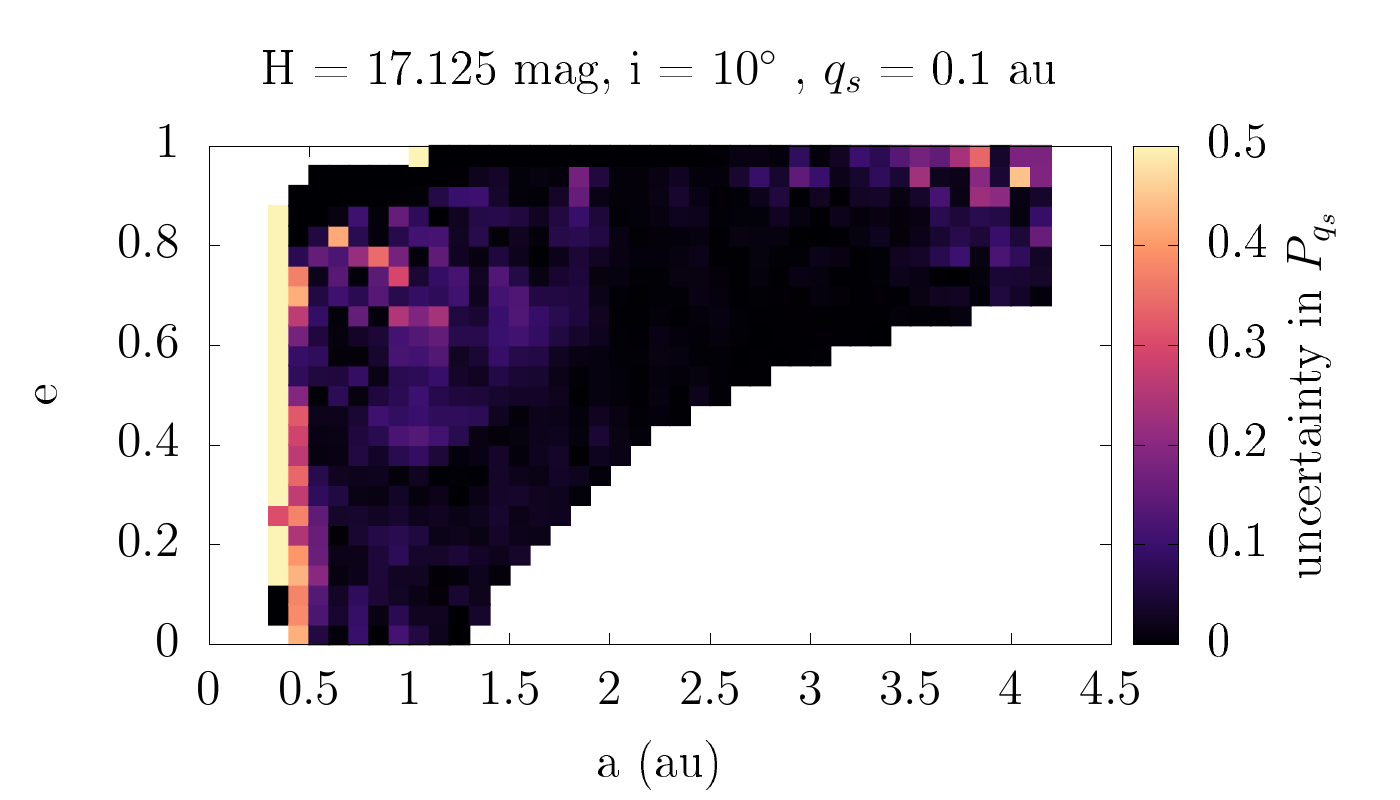}
\caption{Left panels: The distribution $P_{q_s}$ on the ($a,e$) plane that an asteroid with $i=10\deg$ and $H=17.125\hmag$ has had in the past $q<1\au$ (top), $q<0.5\au$ (middle) and $q<0.1\au$ (bottom). Right panels: The respective calculated uncertainties in the same plane and the same slice in the grid. Note the different scale in the color coding for a more comprehensible presentation.}
\label{fig:prob10-17.125}
\end{figure*}

In the left panels of Fig.~\ref{fig:prob10-17.125}, we show the $P_{q_s}$ distribution in the ($a,e$) plane. We have taken a "slice" of the grid with $H=17.125\hmag$ and $i=10\deg$ kept fixed. The reason $i=10\deg$ was chosen was because for this value the estimated number of asteroids with $17<H<18.5\hmag$ is the maximum \citep{Granvik2018}. Three cases with $q_s=1,0.5,0.1\au$ are plotted to cover a large range in $q$, with the color coding corresponding to $P_{q_s}$.

The first obvious characteristic is that the $q=q_s$ limit is marked by the edge of the yellow region (probability equal to unity). This is, as expected, a natural outcome of the fact that these cells already have $q$ below the threshold $q_s$. Obviously, the probability decreases for cells with larger $q$ values. Along the same $q$ line, the probability is higher for cells close to the 3:1 MMR and the $\nu_6$ secular resonance and lower for $a>2.5\au$ due to resonant asteroids undergoing large oscillations in eccentricity, a result which is in accordance with findings of \citet{Marchi2009}. The missing cells at the top left corner of the plots correspond to the location of $q=\bar{q}_*(17.0<H<17.25)$. 

The right panels show the uncertainty in the $q_{min}$ probability calculated with the method described in section~\ref{sec:methods} and refer to the same slice of the ($a,e,i,H$) grid. We notice that the uncertainty increases with decreasing $q_s$ in the edges of the $a$ dimension which are less populated compared to the center of the $a$ range. The yellow cells in the innermost edge, corresponding to uncertainty of unity, suggest that an extremely small number of test asteroids, perhaps just one, reached that cell. Consequently, either the even or the odd population has a probability of unity while the other has zero. 

\begin{figure*}
\centering
\includegraphics[width=0.49\textwidth]{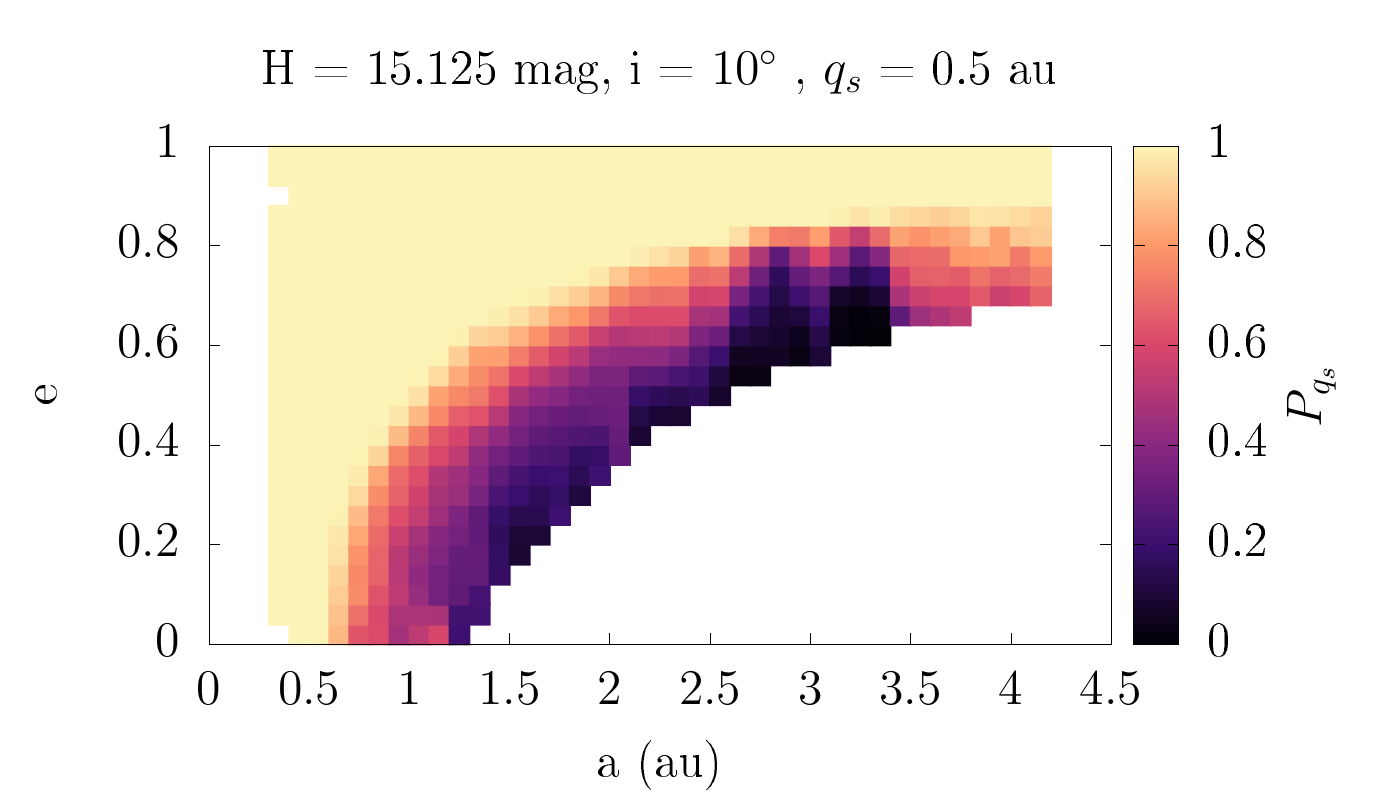}
\includegraphics[width=0.49\textwidth]{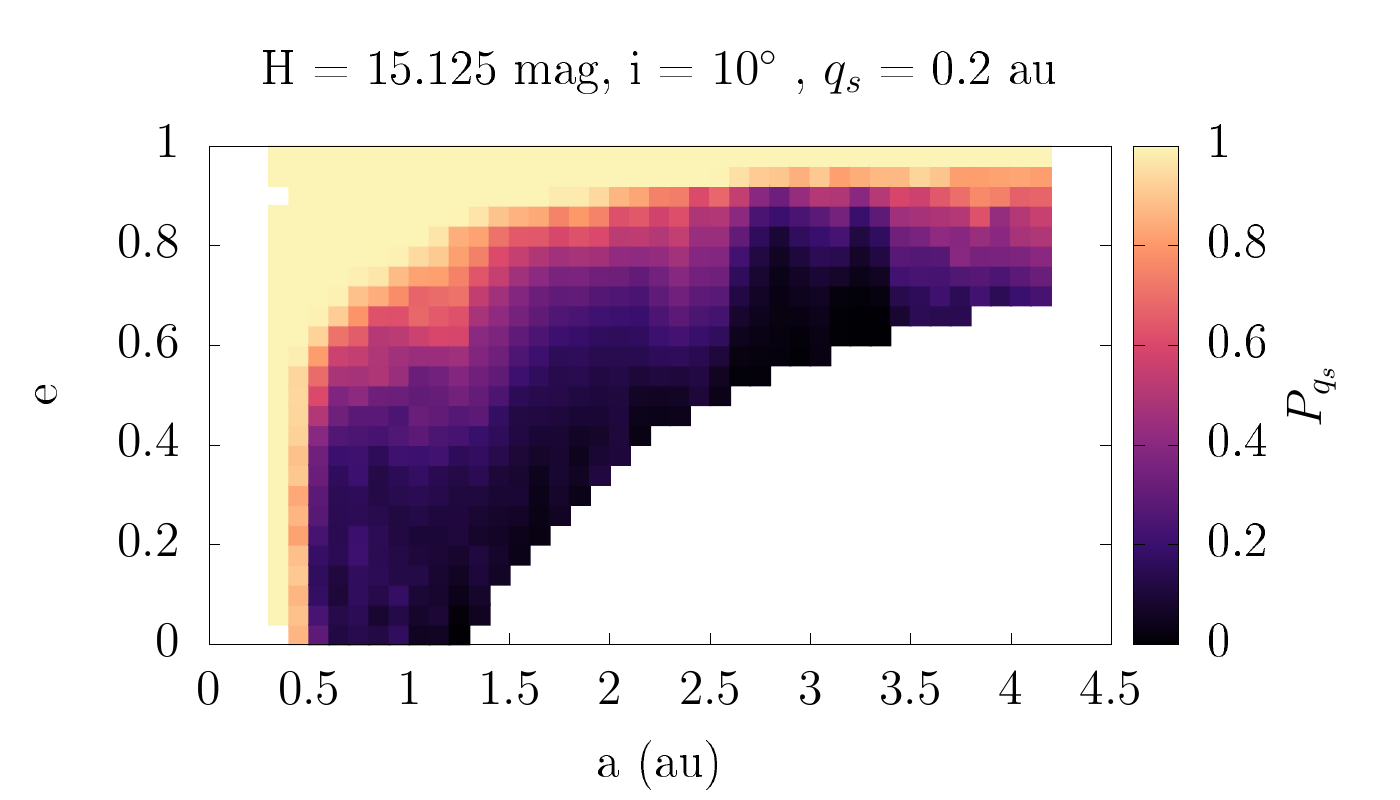}

\includegraphics[width=0.49\textwidth]{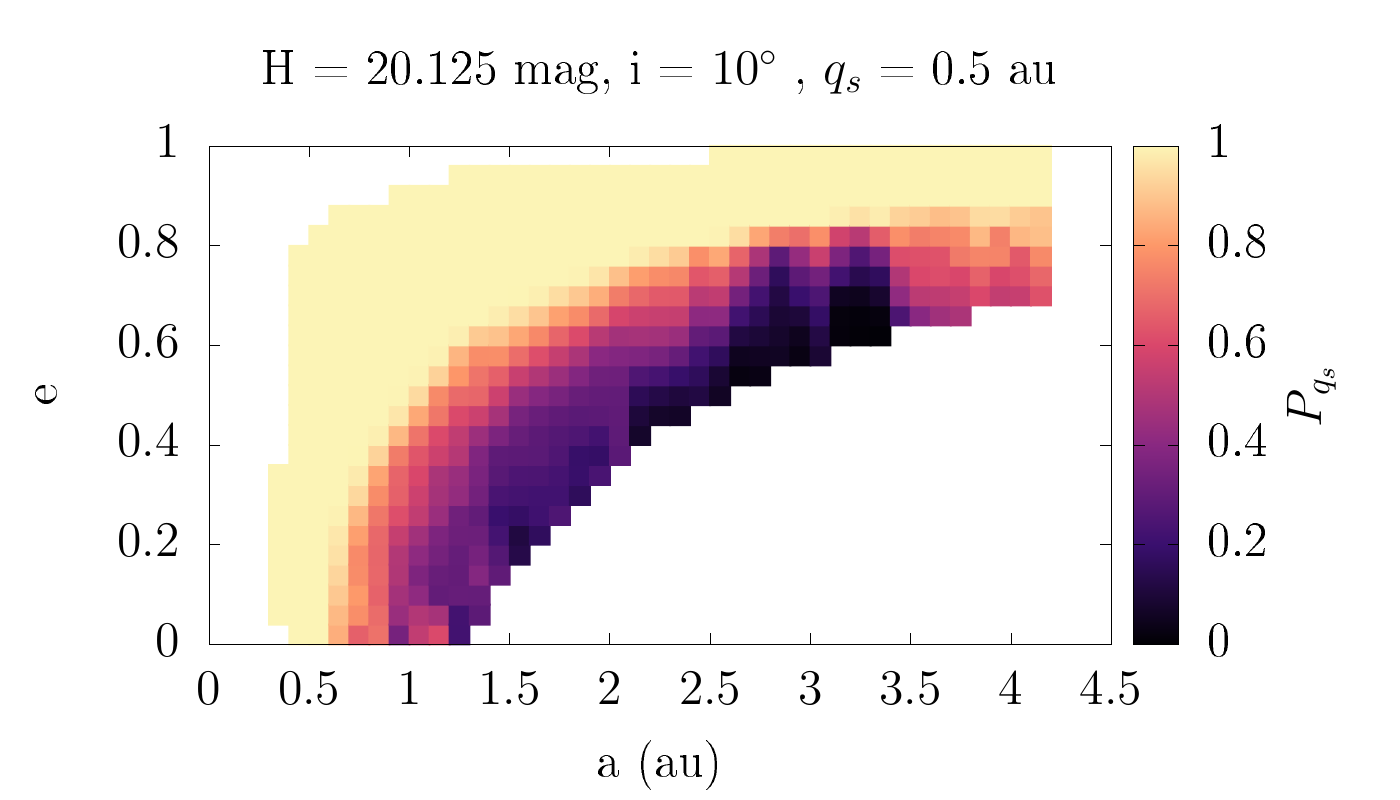}
\includegraphics[width=0.49\textwidth]{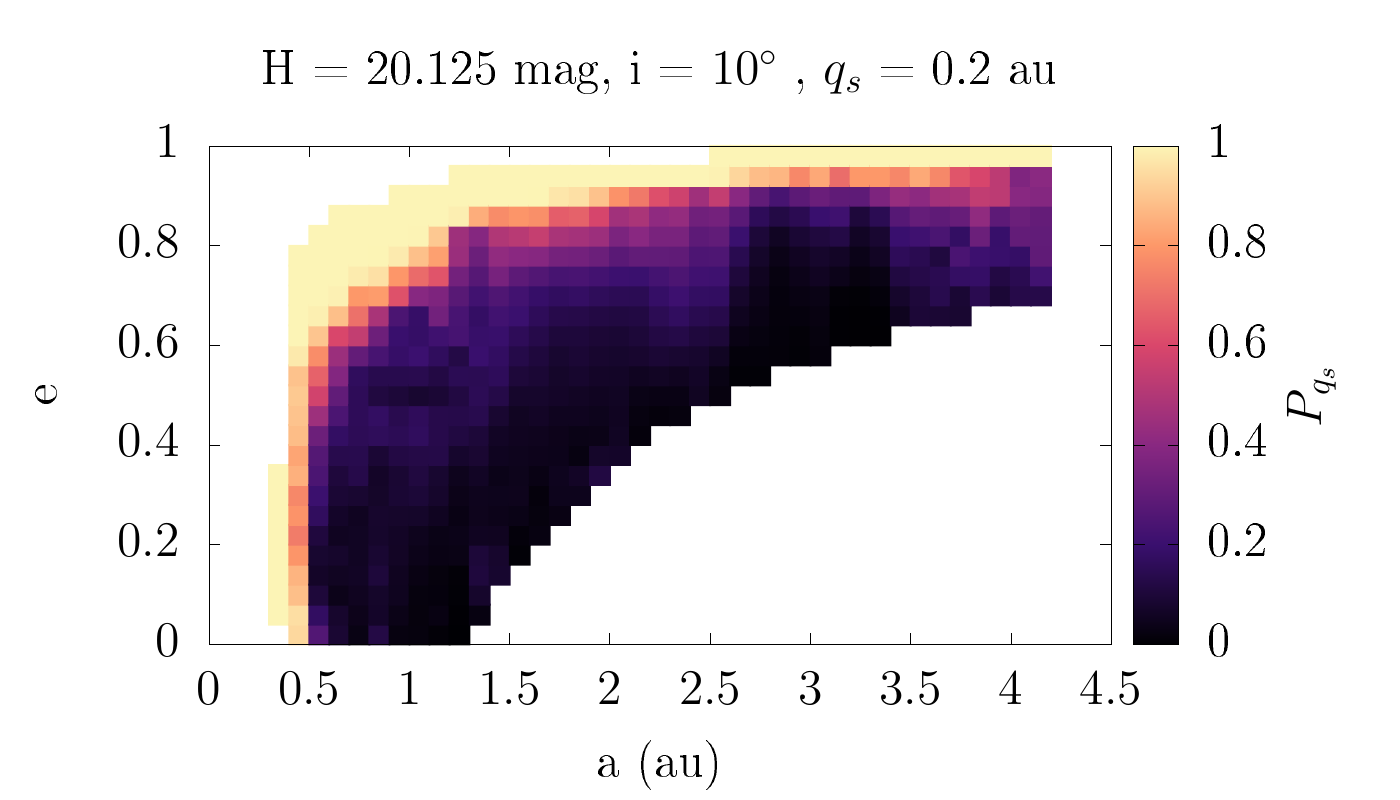}

\includegraphics[width=0.49\textwidth]{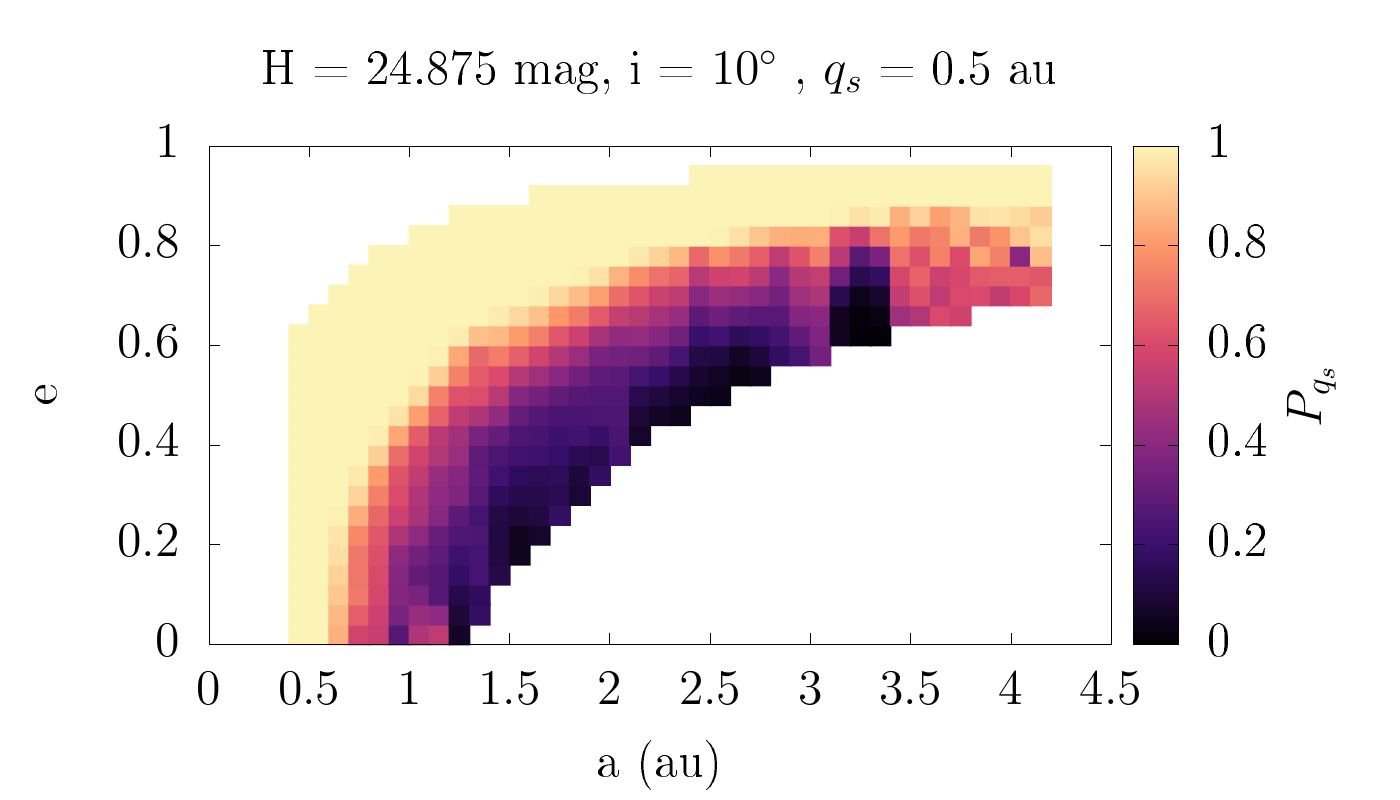}
\includegraphics[width=0.49\textwidth]{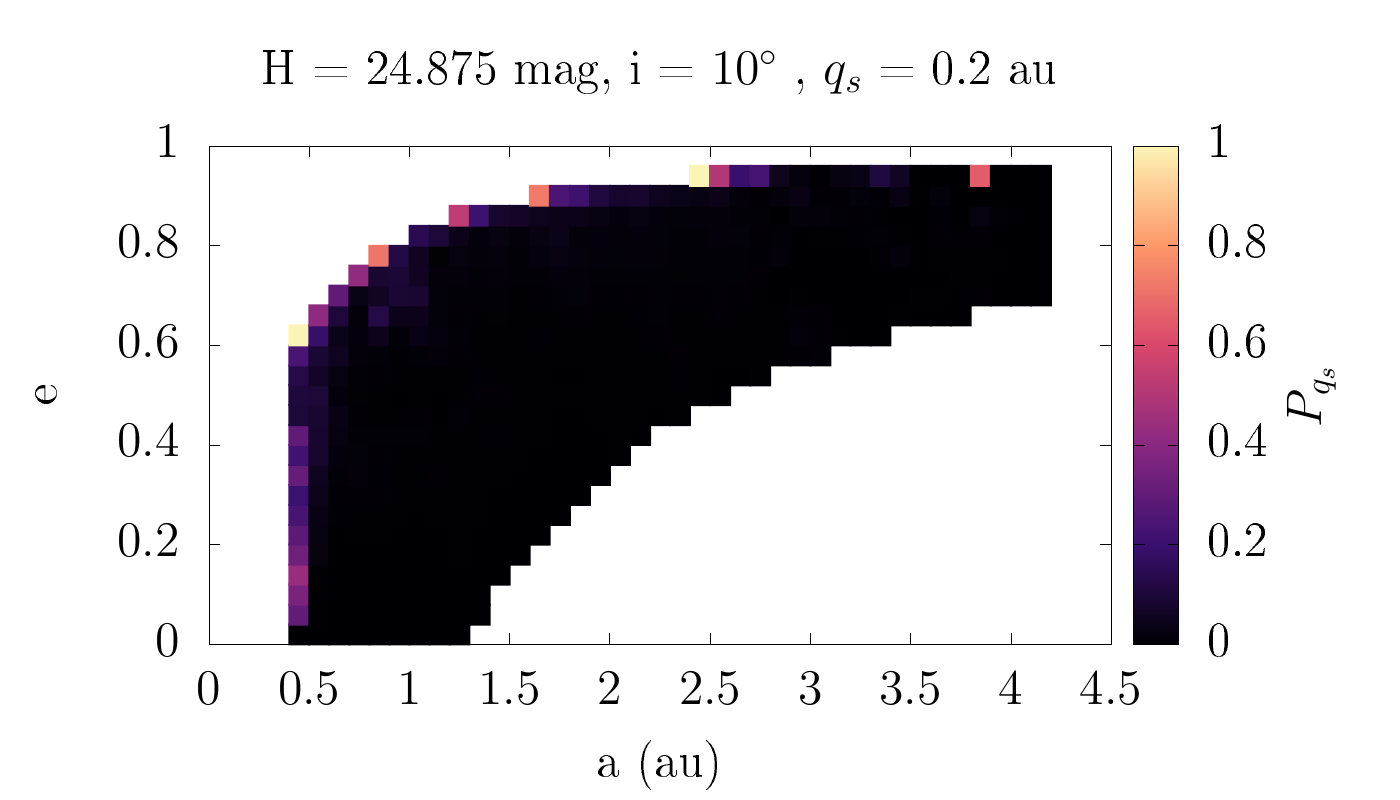}

\caption{The distribution of $P_{q_s}$ on the ($a,e$) plane that an asteroid with $i=10\deg$ has had in the past $q<0.5\au$ (left panels) and $q<0.2\au$ (right panels) for $H=15.125\hmag$ (top),  $H=20.125\hmag$ (middle) and  $H=24.875\hmag$ (bottom).}
\label{fig:prob_h-10}
\end{figure*}

The variation of the $P_{q_s}\left(a,e,i,H\right)$ distributions with $H$ can be seen in Fig.~\ref{fig:prob_h-10} for $q_s=0.5\au$ (left panels) and $q_s=0.2\au$ (right panels). For $q_s=0.5\au$, it is apparent that there are small differences in the distribution of $P_{q_s}$, but the most prominent feature is the absence of cells for $q<\bar{q}_*(H)$, due to the disruption of asteroids at small $q$ that we included in our model. The main factor that affects the distribution of probabilities with varying $H$ is $\beta_\text{ER}(a,e,i,H)$. For an overview of the average source specific variation of $\beta$ with $H$ see fig.~13 in \citet{Granvik2018}. For $q_s<0.2\au$ we notice that typically $P_{q_s}\left(a,e,i,H\right)$ decreases monotonously with increasing $H$. That is a consequence of the disruption of asteroids at $q=\bar{q}_*(H)$. By considering the test asteroid totally destroyed after its $q$ takes the critical value and disregarding any future evolution, we fill less $q$ counters that are closer to the Sun for smaller asteroids compared to larger asteroids. This trend however breaks for cells that are away from strong resonances and asteroids that reside there have low probabilities of having acquired small $q$ values. For larger $q_s$ the trend also disappears. 

\begin{figure*}
\centering
\includegraphics[width=0.49\textwidth]{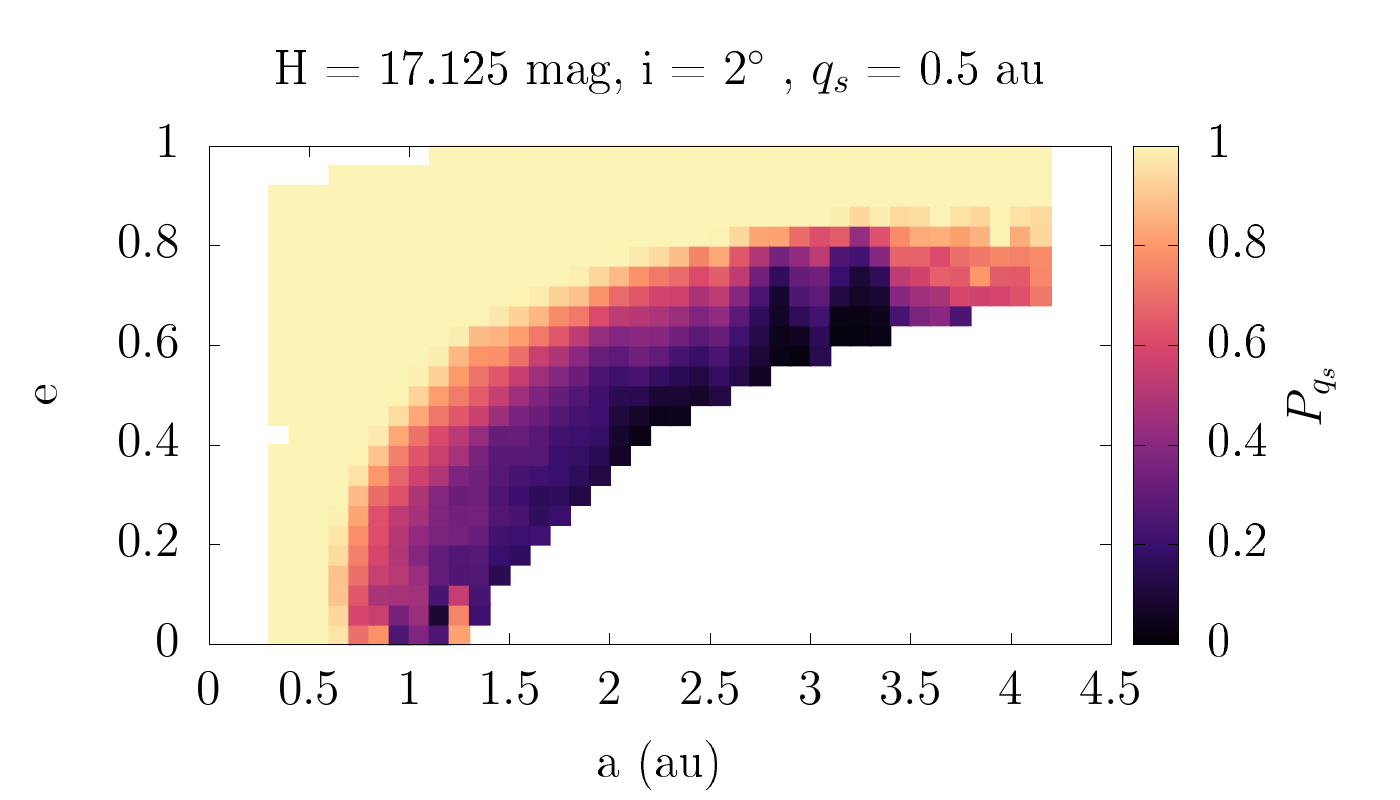}
\includegraphics[width=0.49\textwidth]{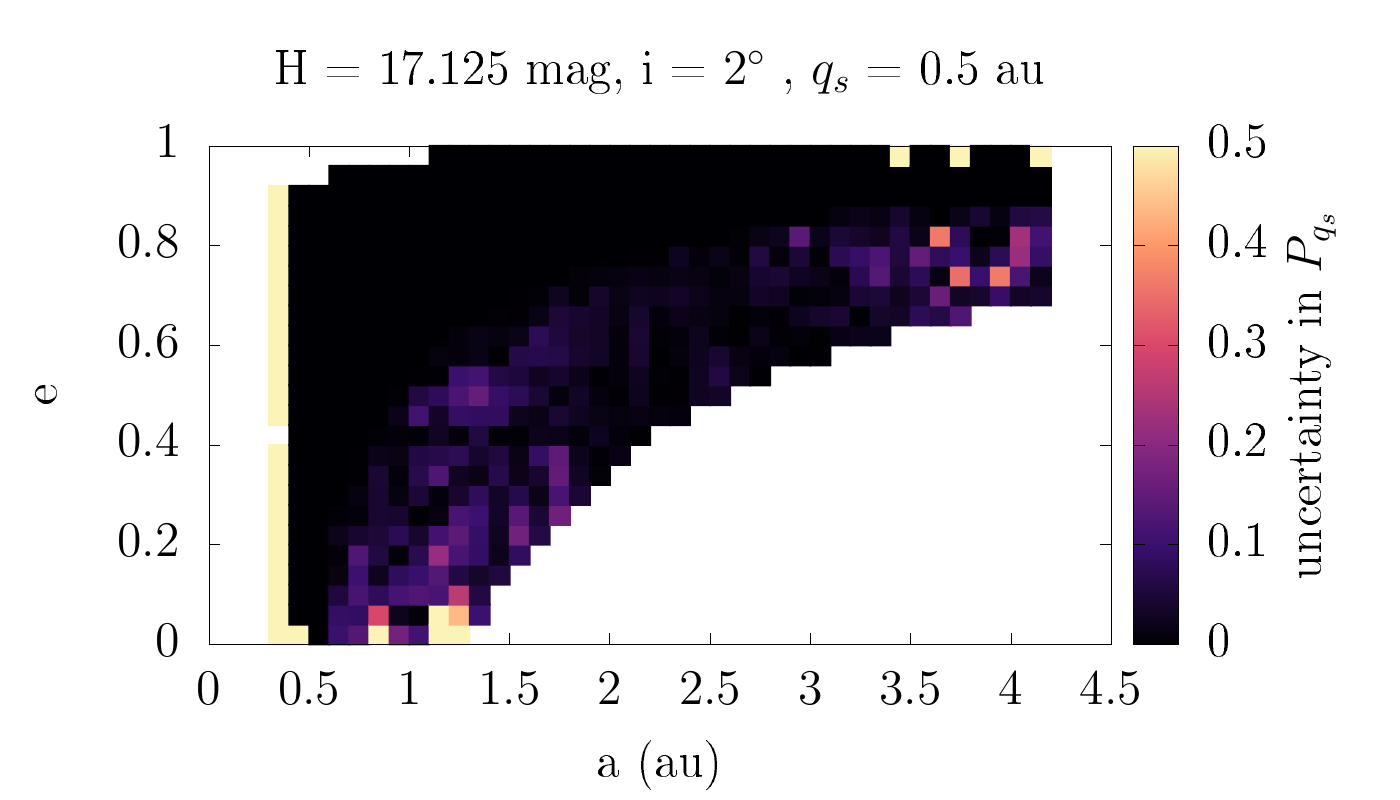}

\includegraphics[width=0.49\textwidth]{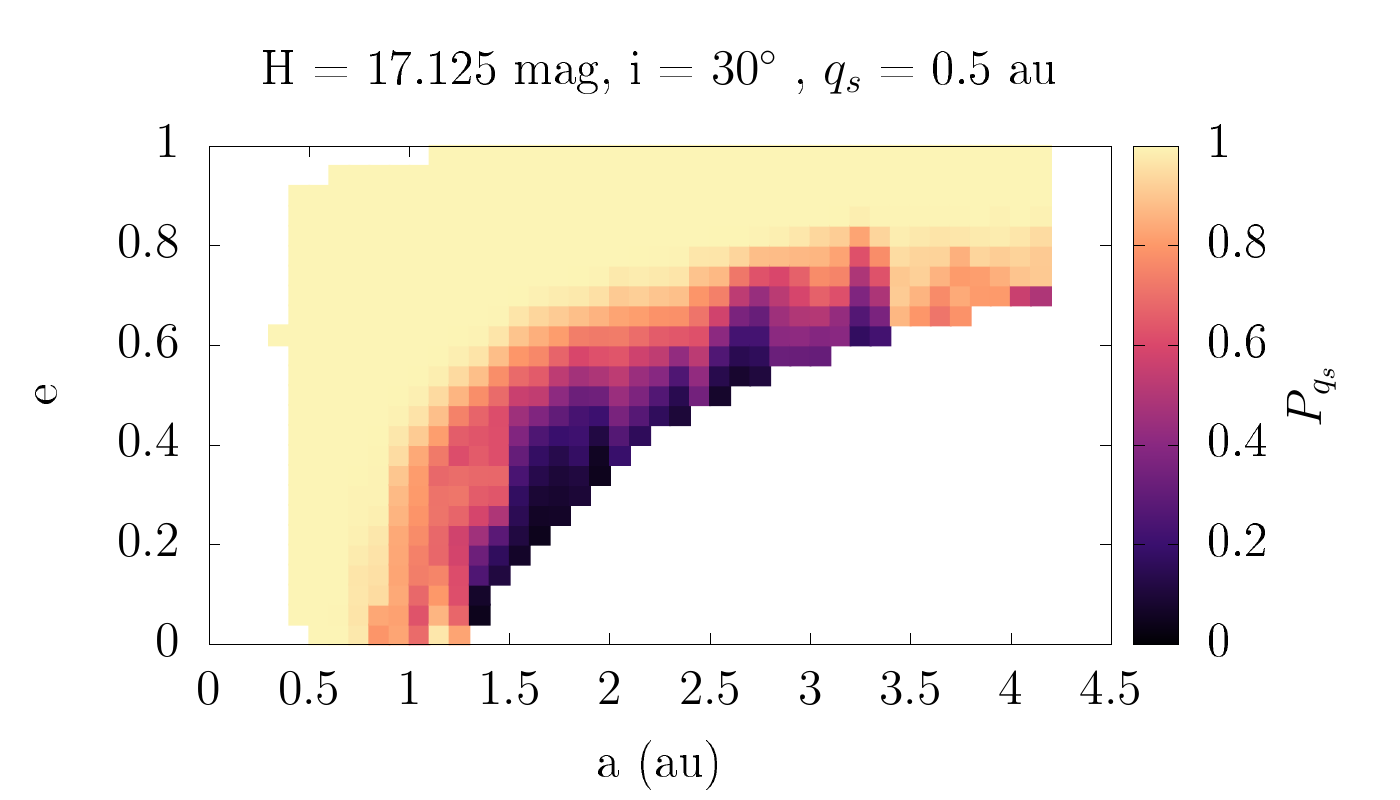}
\includegraphics[width=0.49\textwidth]{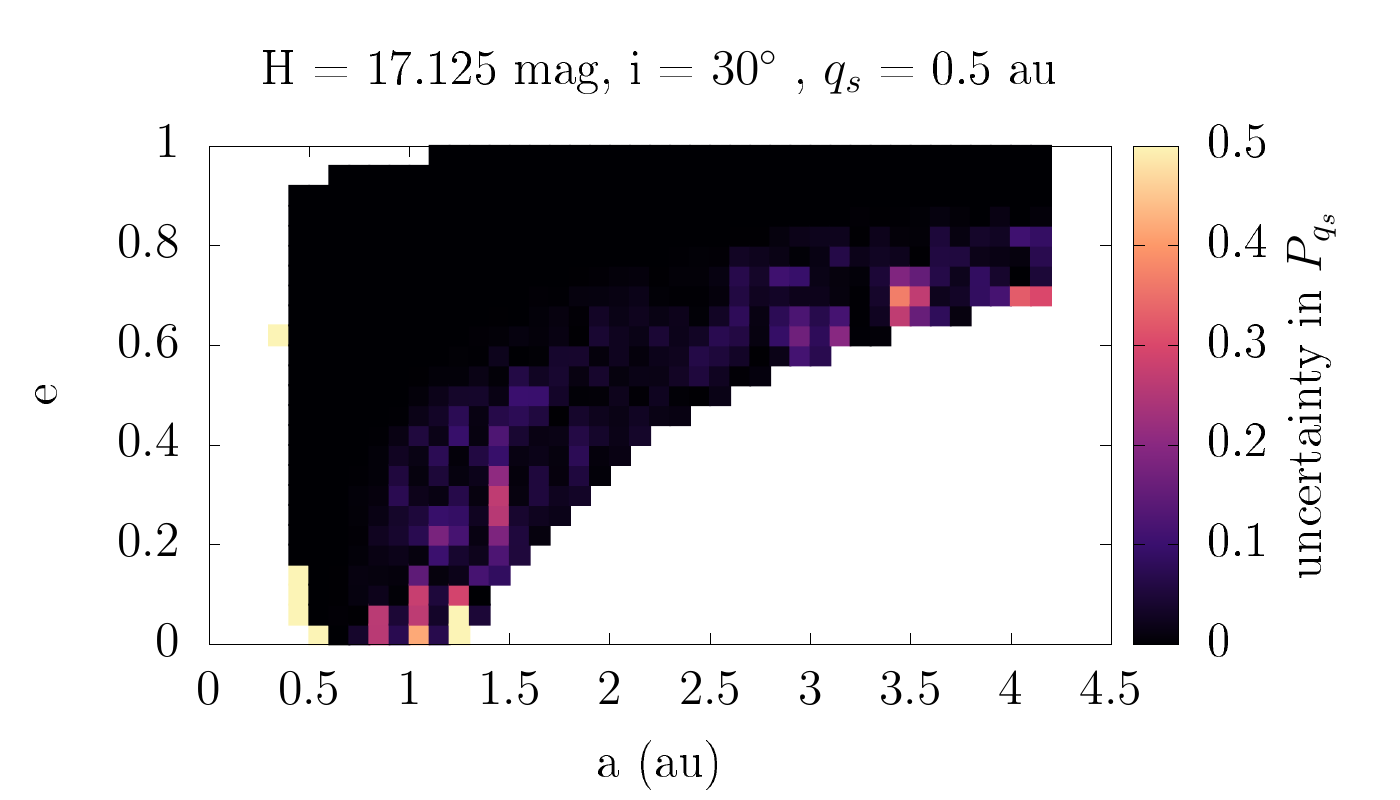}

\includegraphics[width=0.49\textwidth]{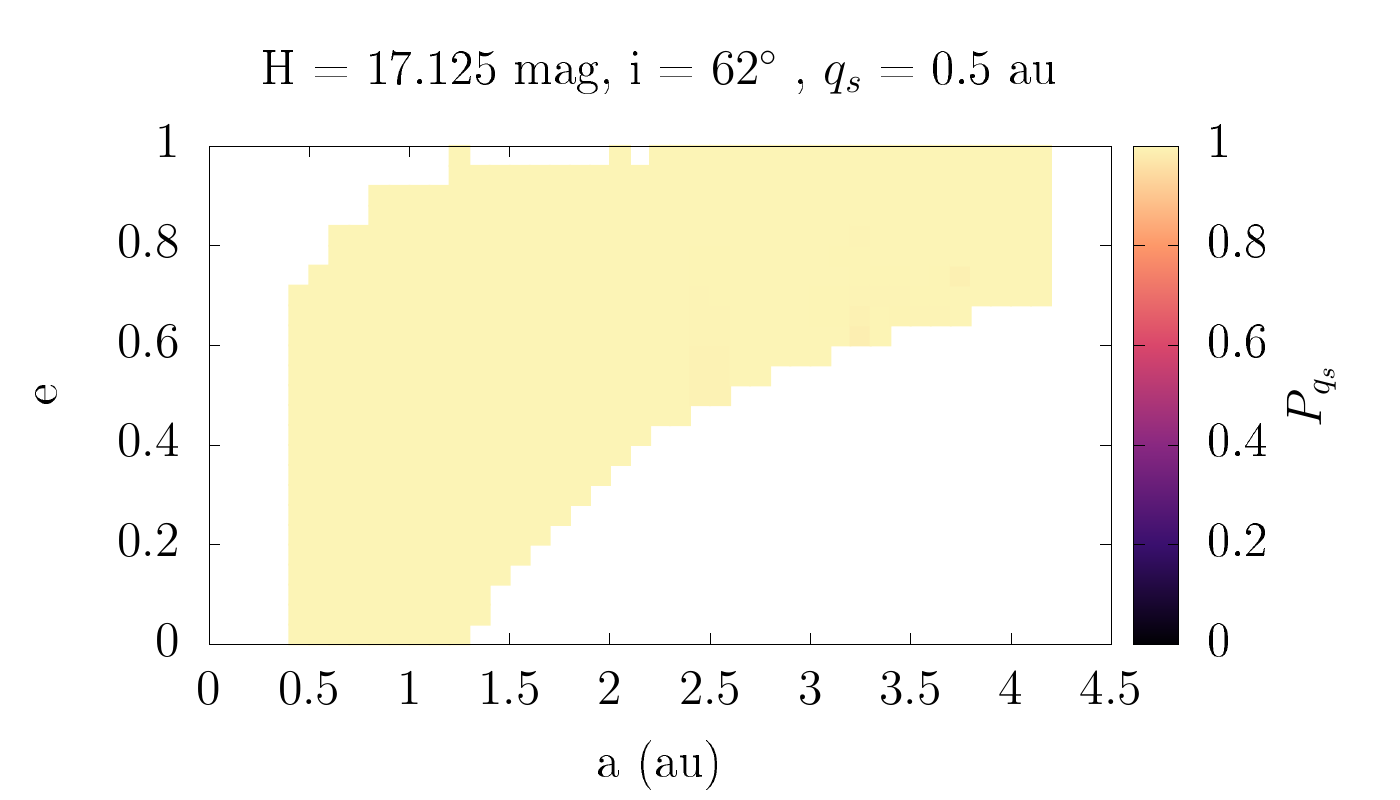}
\includegraphics[width=0.49\textwidth]{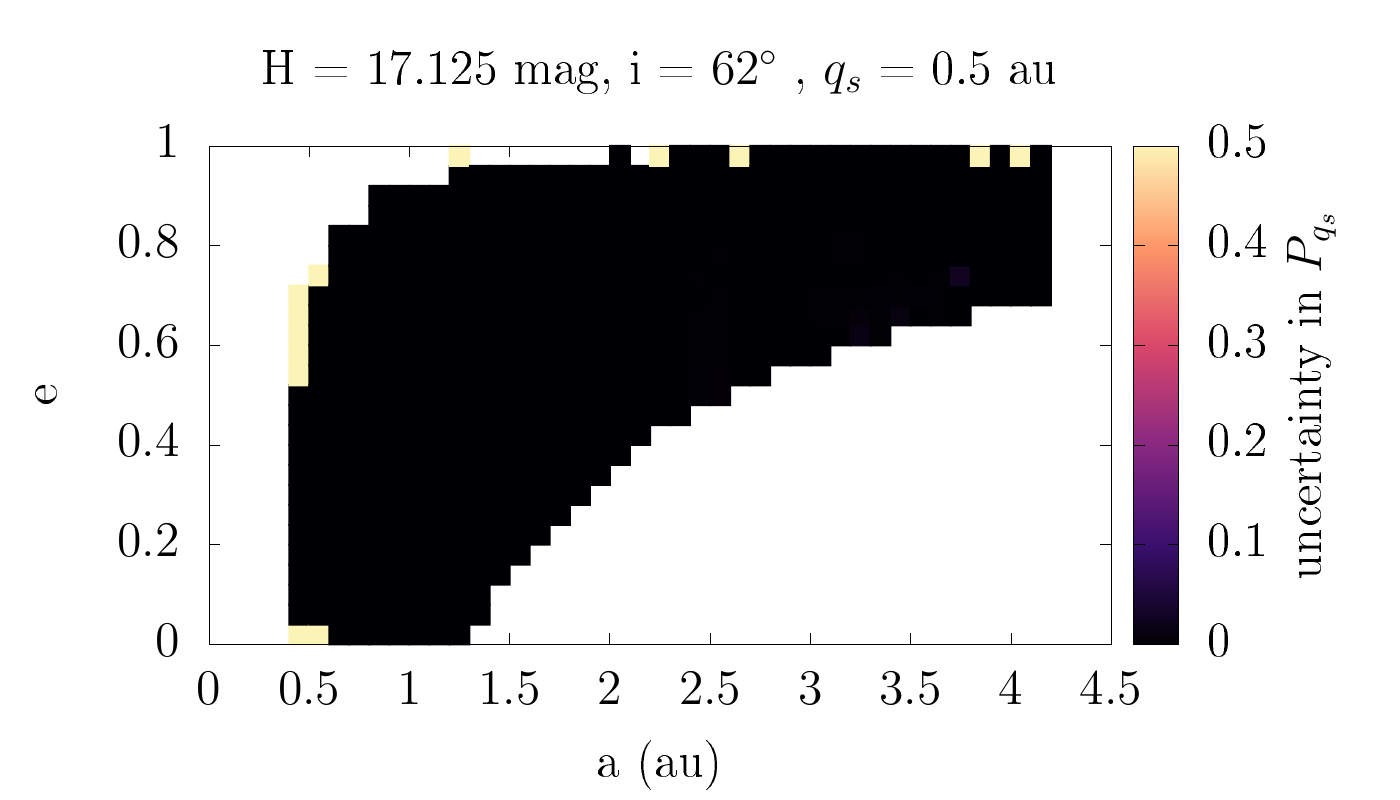}
\caption{The ($a,e$) distribution of $P_{q_s}$ and respective uncertainty for asteroids with $H=17.125\hmag$ and $i=2\deg$ (top panel), $i=30\deg$ (middle panel) and $i=62\deg$ (bottom panel).}
\label{fig:prob_inc-17.125}
\end{figure*}

In Fig.~\ref{fig:prob_inc-17.125}, we show that there can be significant changes in $P_{q_s}\left(a,e,i,H\right)$ for different $i$ values. The left panels plot the probabilities distribution for two slices in the grid: $i=2\deg$ and $30\deg$ with $H=17.125\hmag$. Focusing on $q_s=0.5\au$, we find that the calculated probabilities increase with increasing $i$. This is not surprising because asteroids that reach this large $i$ values are subject to the Lidov--Kozai mechanism \citep{Zeipel1910, Lidov1962, Kozai1962}. These asteroids undergo large, coupled oscillations in $i$ and $e$, while the argument of perihelion ($\omega$) can either circulate, with a precession frequency correlated to the oscillation of $e$ and $i$, or librate around $90\deg$ or $270\deg$ (for a review, see \citet{Takashi&Katsuhito2019,Morby2002}). Consequently, it is \textit{certain} that in the past their $q$ has reached below $0.5\au$. Hence, the uncertainties are zero.

\subsection{Dwell times}

\begin{figure*}
\centering
\includegraphics[width=0.49\textwidth]{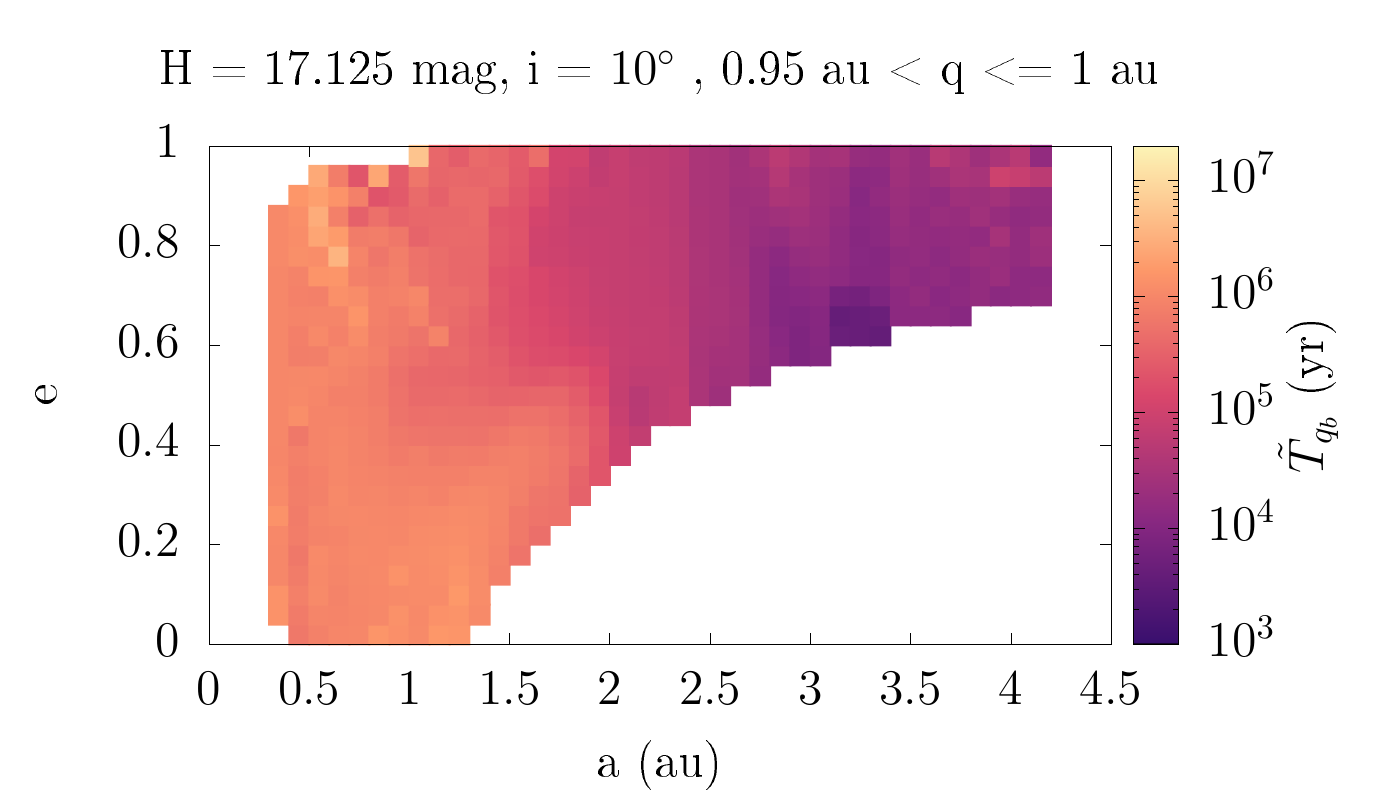}
\includegraphics[width=0.49\textwidth]{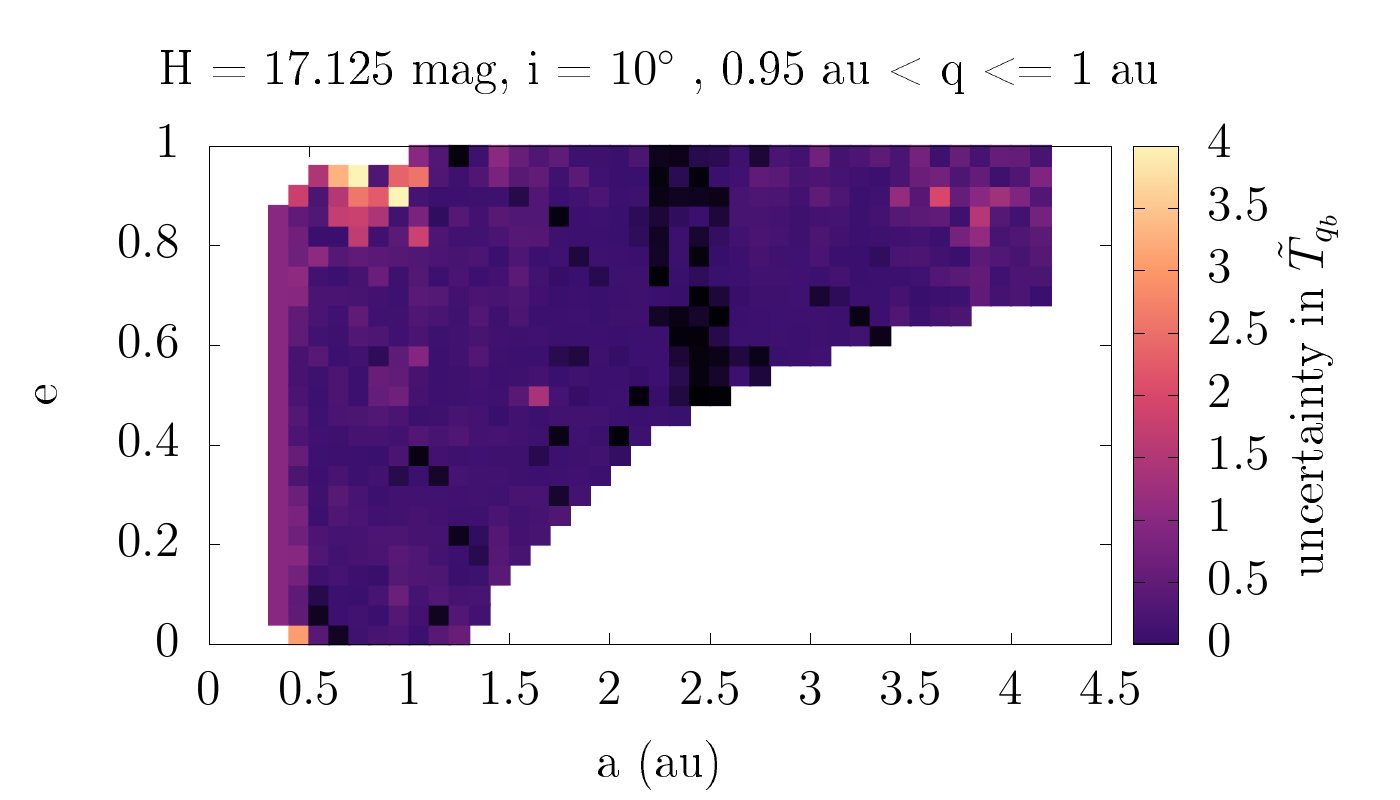}

\includegraphics[width=0.49\textwidth]{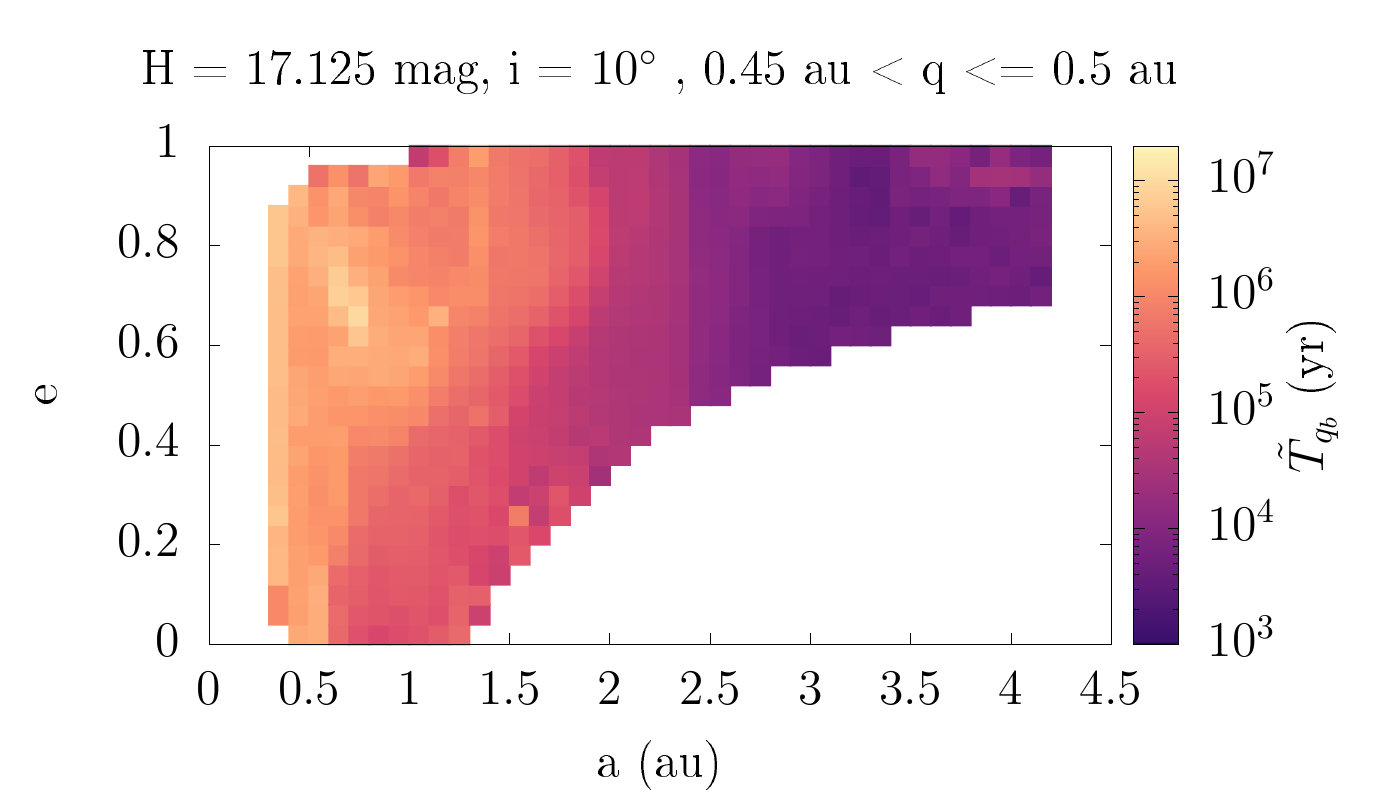}
\includegraphics[width=0.49\textwidth]{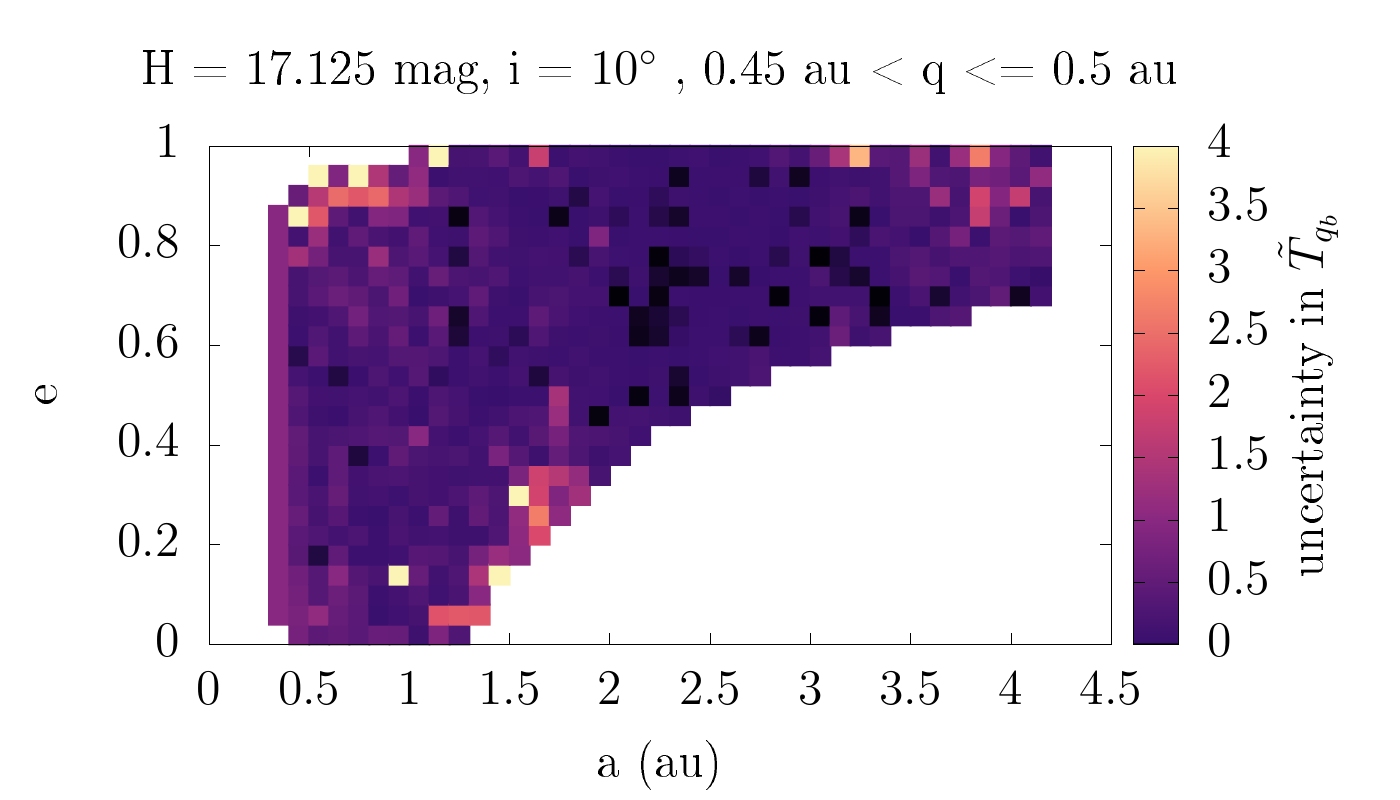}

\includegraphics[width=0.49\textwidth]{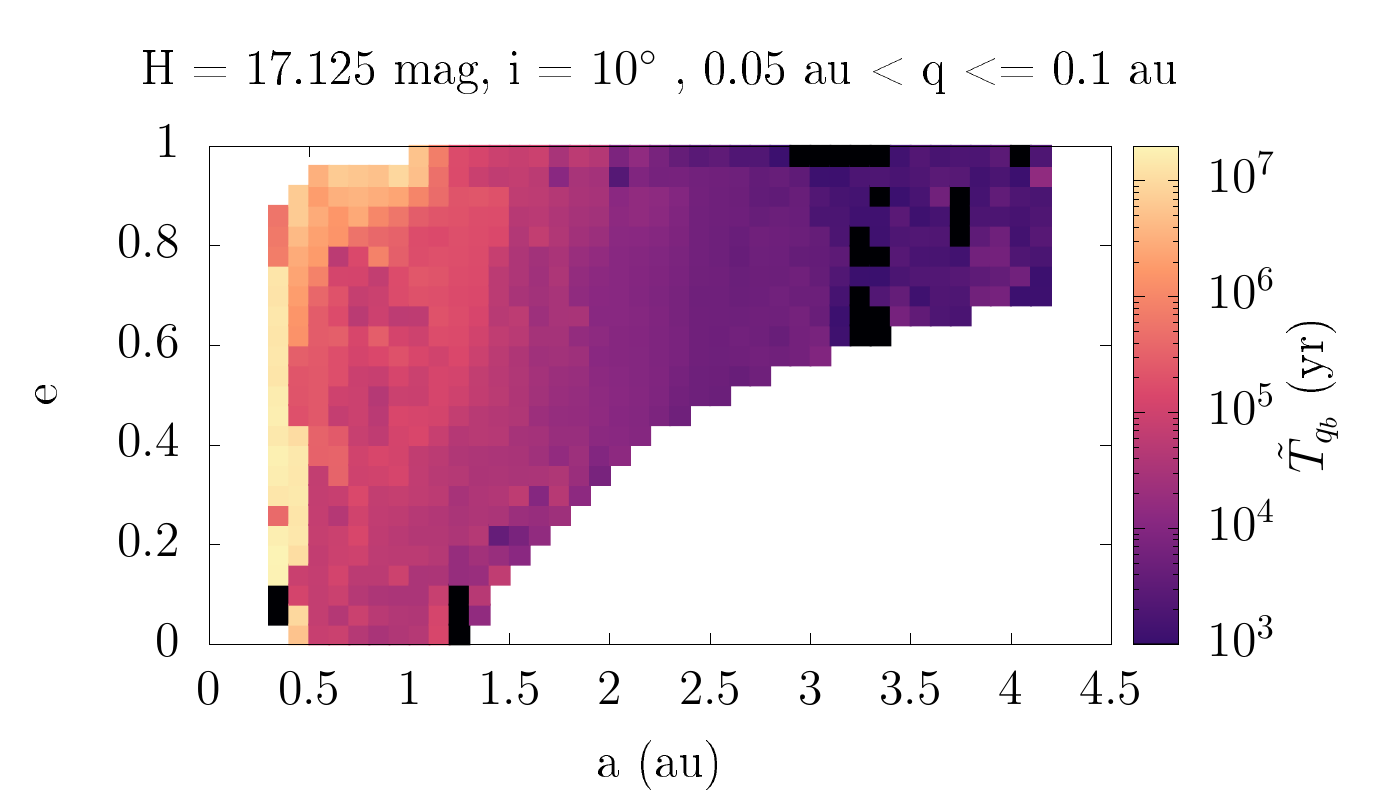}
\includegraphics[width=0.49\textwidth]{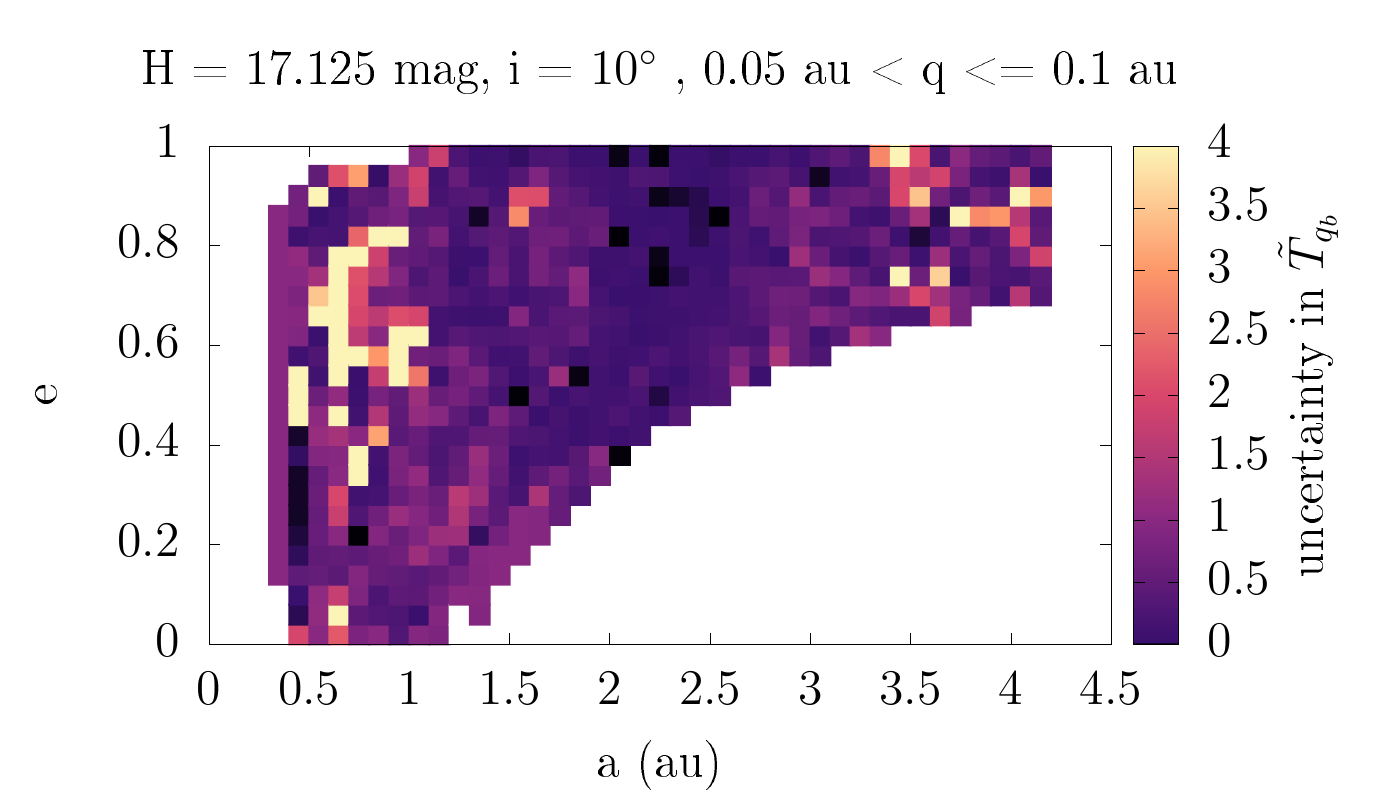}
\caption{Left panels: The ($a,e$) distribution (for $i=10\deg$ and $H=17.125\hmag$) of the median values of the total time, recorded in these cells, that an asteroid has had spent having $q$ in the range $0.95<q\leq 1\au$ (top), $0.45<q\leq 0.5\au$ (middle) and $0.05<q\leq 0.1\au$ (bottom). Right panels: The respective calculated uncertainties in the same plane and slice of the grid. }
\label{fig:dwell10-17.125}
\end{figure*}

In Fig.~\ref{fig:dwell10-17.125}, we show the ($a,e$) distribution of the median dwell times and their uncertainties in the bins with $0.05<q\leq 0.1$, $0.45<q\leq 0.5$ and $0.95<q\leq 1\au$ for asteroids with $H=17.125\hmag$ and $i=10\deg$. 

Even for the smallest $q$ range considered, the dwell times can be very large, even in the order of $10^6$~yr. The dwell times become significantly shorter for large $H$ because we account for the destruction of asteroids at small distances from the Sun and this affects small asteroids (that is, those with large $H$) more than the large ones. On the other hand, the dwell times are in general smaller for larger $a$ values, exterior to the 3:1 MMR, because that region is densely populated by high-order MMRs with Jupiter thus making it fairly unstable.  
 
\begin{figure*}
\centering
\includegraphics[width=0.49\textwidth]{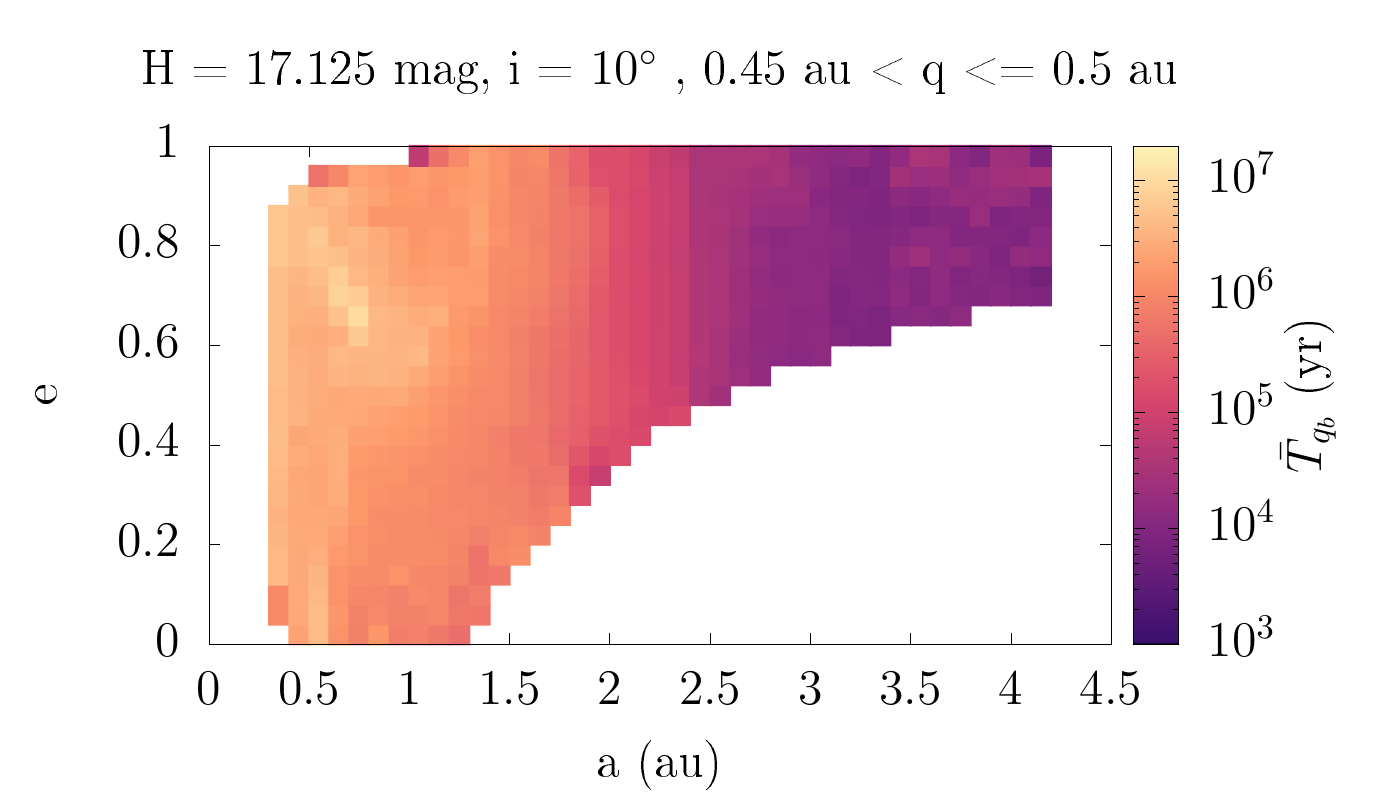}
\includegraphics[width=0.49\textwidth]{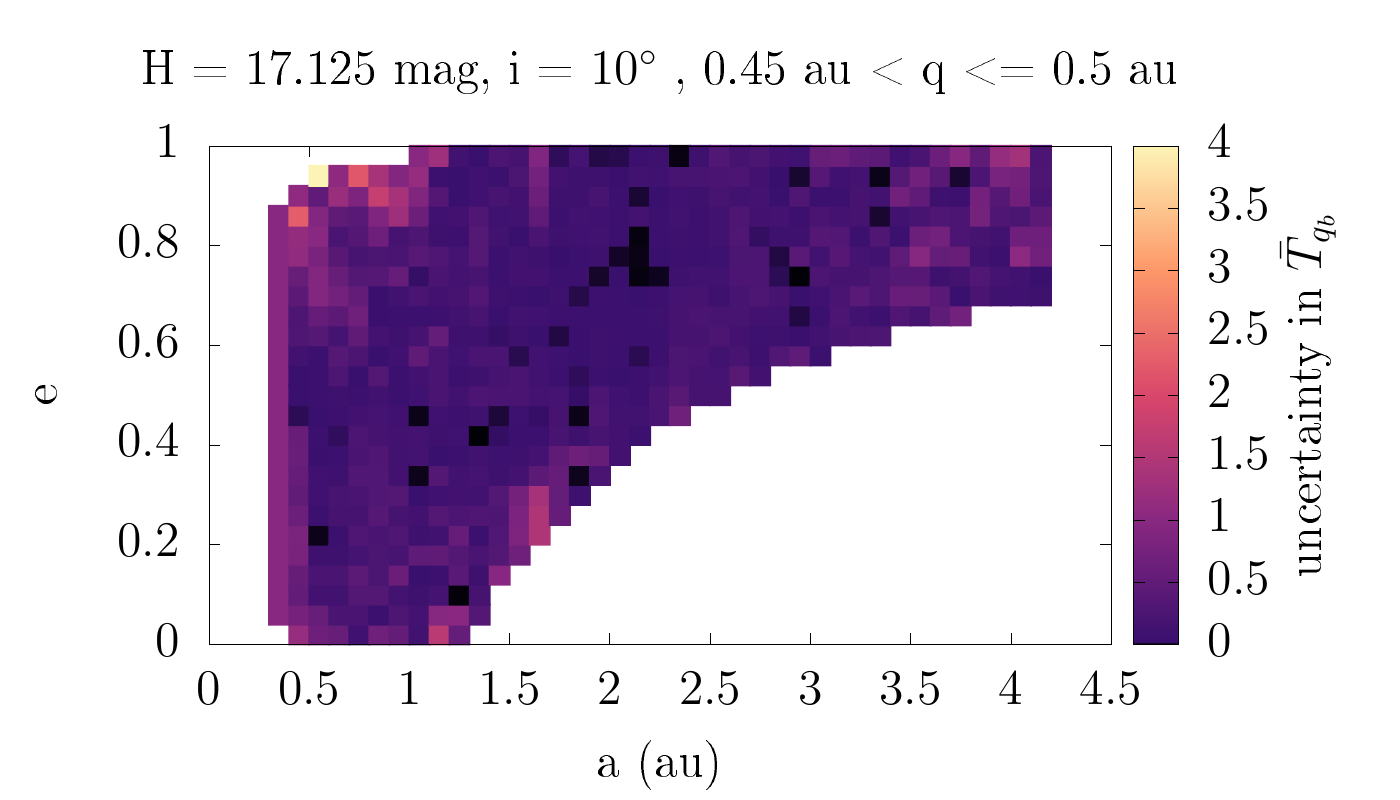}
\includegraphics[width=0.49\textwidth]{dwell_median0.5.pdf}
\includegraphics[width=0.49\textwidth]{unc_median0.5.pdf}
\caption{A comparison between the ($a,e$) distributions of the mean (top panels) and median (bottom panels) dwell times of the perihelion distance of an asteroid in the range $0.45<q\leq 0.5\au$ for $i=10\deg$ and $H=17.125\hmag$. The nominal values are on the left and the corresponding uncertainties are on the right. }
\label{fig:avg-med10-17.125}
\end{figure*}

\begin{figure}
\centering
\includegraphics[width=0.48\textwidth]{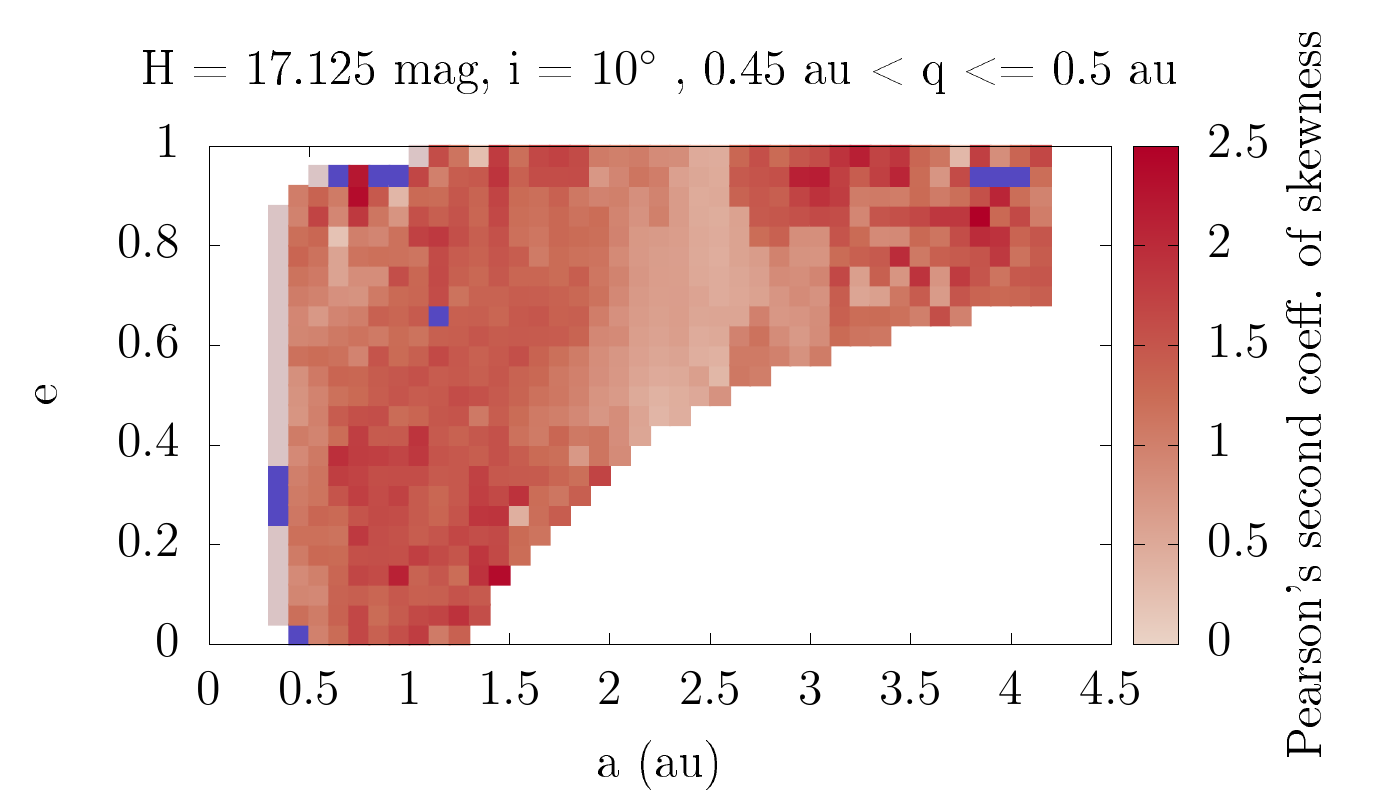}

\includegraphics[width=0.42\textwidth]{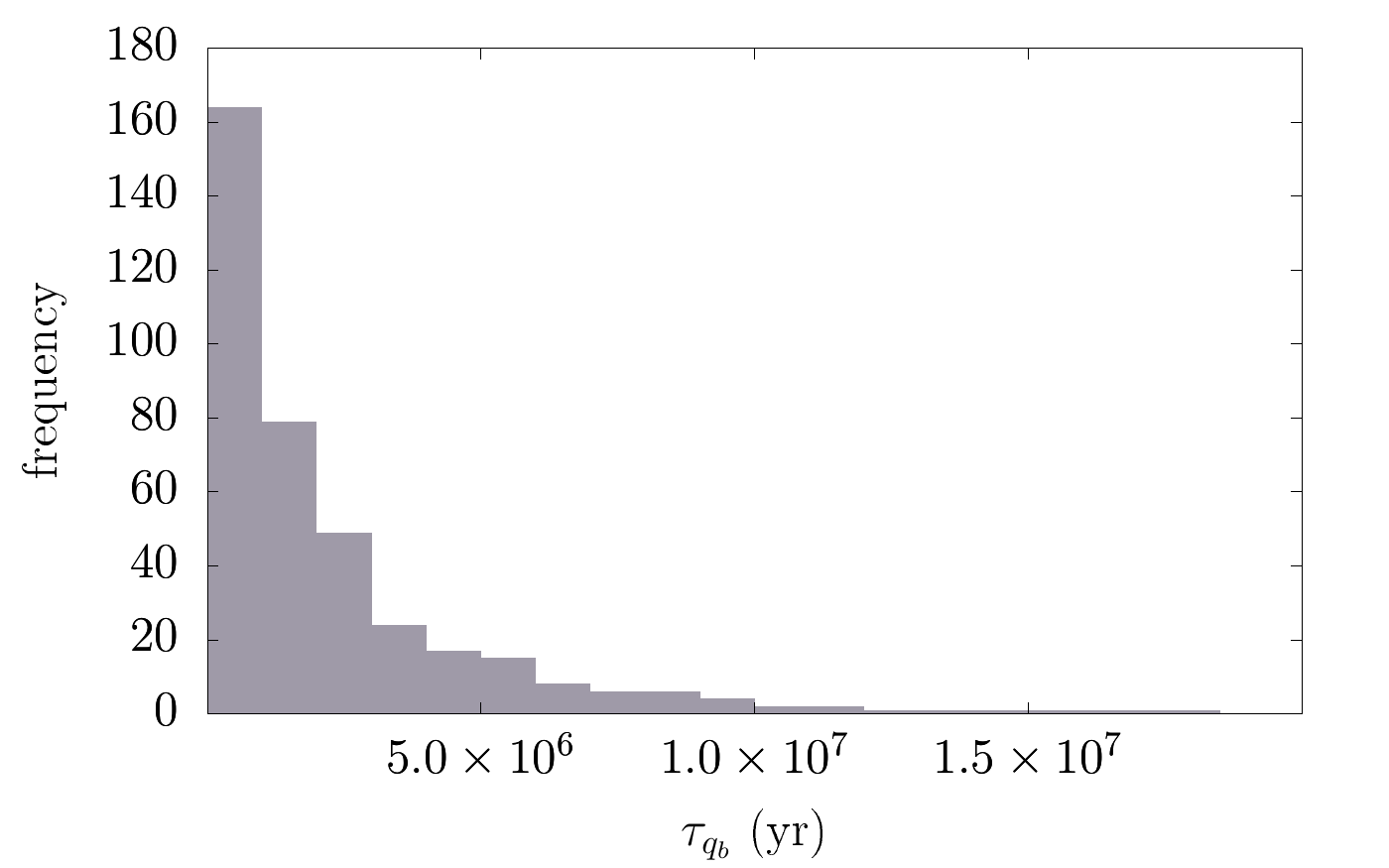}
\caption{Top panel: The distribution of Pearson's second coefficient of skewness of the recorded dwell times in the ($a,e$) plane for cells with $i=10\deg$ and $H=17.125\hmag$ concerning the $0.45\au<q<0.5\au$ bin. The blue points correspond to cells that have negative skewness. Bottom panel: A histogram of the $\tau_{q_b}$, recorded for a cell with $a=0.95\au$, $e=0.5$, $i=10\deg$, $H=17.125\hmag$ and $0.45<q\leq0.5\au$. This is the unweighted raw data coming from all ERs, showing a tail of some very long dwell times.} 
\label{fig:skewness}
\end{figure}

Comparing the mean and median, we find that, in general, mean values are larger than the median, however the differences are not very large (Fig.~\ref{fig:avg-med10-17.125}). In order to estimate how far from a normal distribution is the distribution  of the recorded dwell times within a ($a,e,i,H$) cell, we calculated the skewness of the distribution of the whole sample of dwell time measurements in that cell, coming from every test asteroid that entered it. We used Pearson's second coefficient of skewness, defined as: 
\begin{equation}
sk=\frac{3\left(\text{Mean}-\text{Median}\right)}{\sigma},
\end{equation}
where $\sigma$ is the standard deviation of the weighted distribution of dwell times, and equal to the square root of the variance given by: 
\begin{equation}
\sigma=\sqrt{\sum_{\text{ER}=1}^{6}\beta_{\text{ER}}^2\sigma_{\text{ER}}^2},
\end{equation}
since the measurements are not correlated. In Fig.~\ref{fig:skewness} (top panel), we show the ($a,e$) distribution of the skewness of the dwell times measurements for asteroids with $i=10\deg$ and $H=17.125\hmag$ in the $q$ range $0.45\au<q<0.5\au$. We found the skewness to be positive almost everywhere, with the exception of a few not very frequently-visited cells. This means that there is a tail of some very long dwell times that have been recorded for these cells. We noticed a similar trend for most cells our grid. An example is shown in the bottom panel of Fig.~\ref{fig:skewness}, with the histogram of the raw data of the recorded $\tau_{q_b}$ for a cell with $a=0.95\au$, $e=0.5$, $i=10\deg$, $H=17.125\hmag$ and $0.45<q\leq0.5\au$, coming from all ERs. The width of each bin is $10^6\yr$. 

\subsection{Minimum perihelion distances for objects from specific escape regions}

\begin{figure*}
\centering
\includegraphics[width=0.49\textwidth]{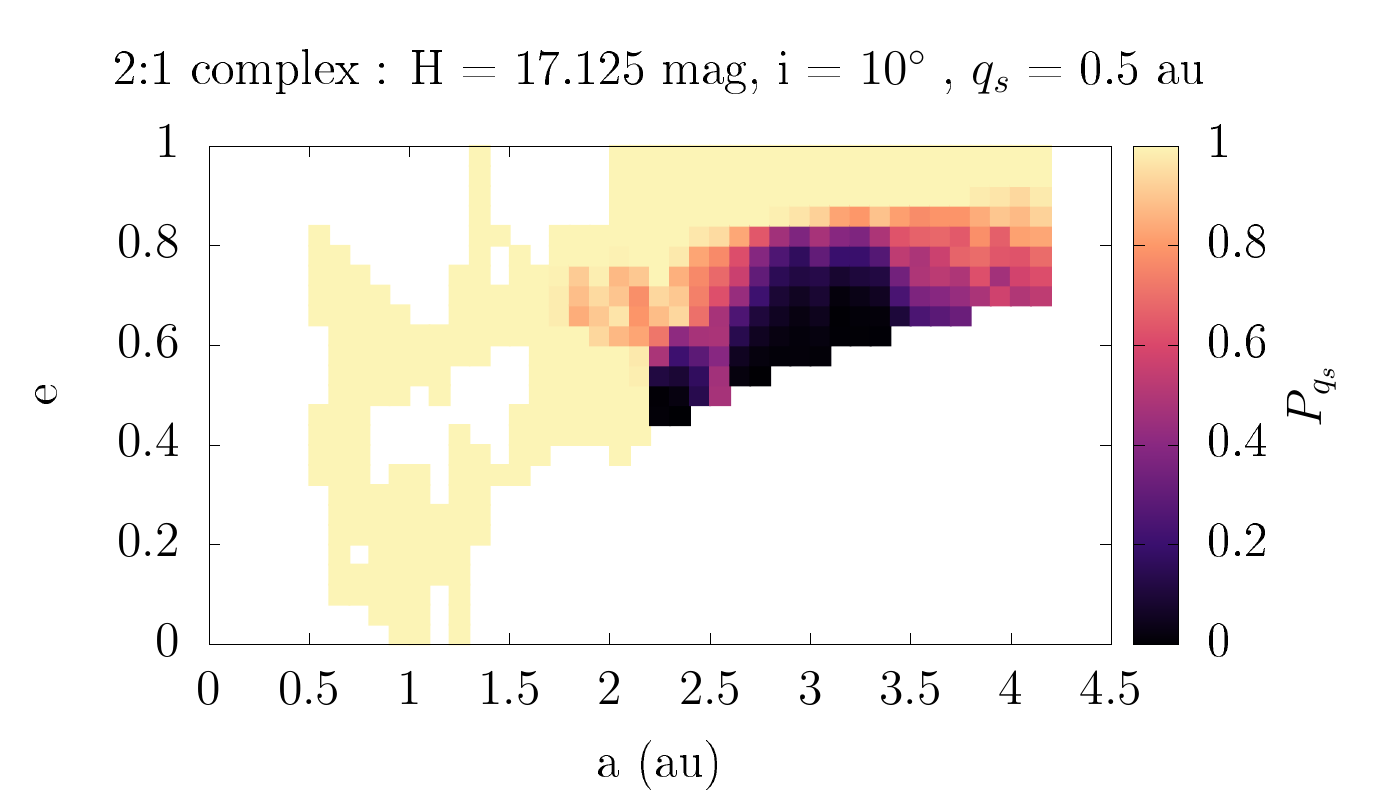}
\includegraphics[width=0.49\textwidth]{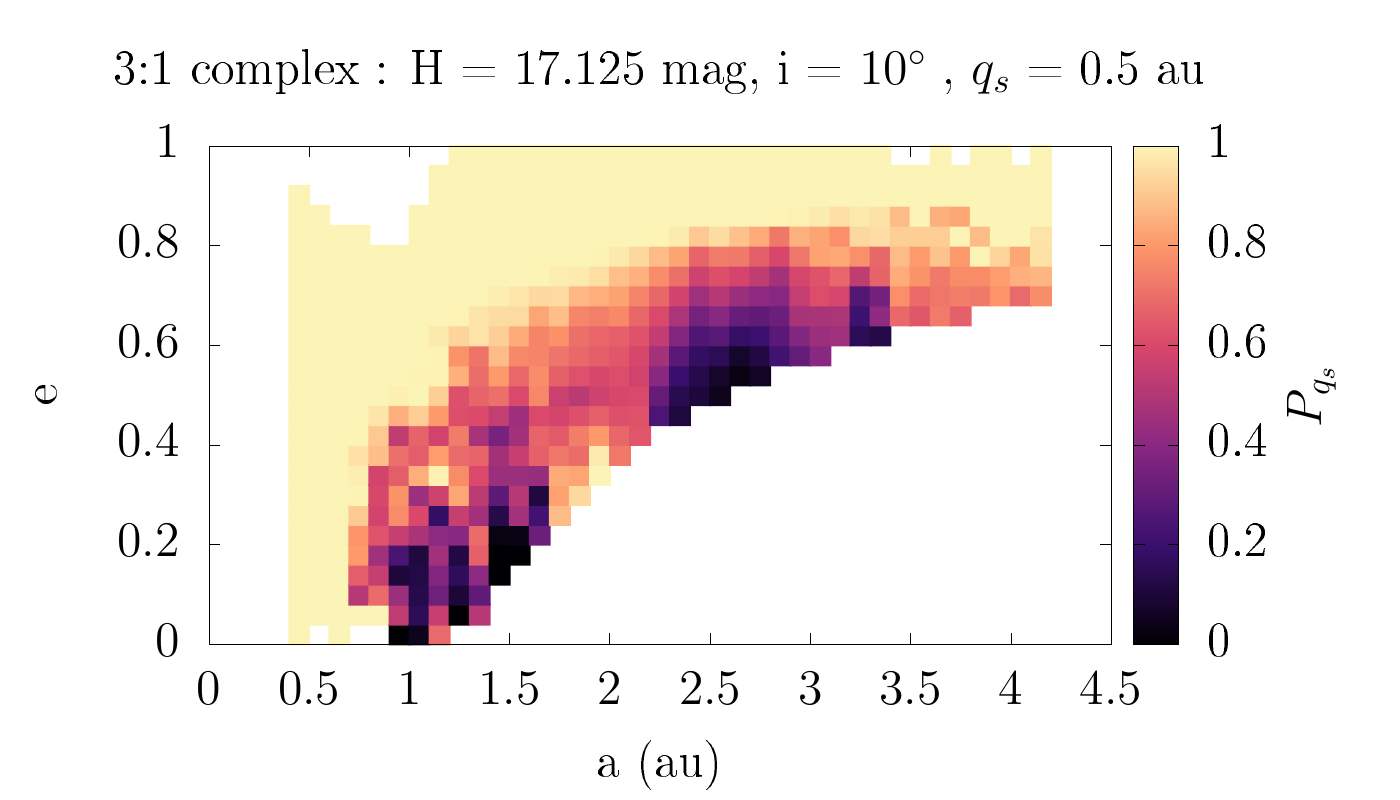}

\includegraphics[width=0.49\textwidth]{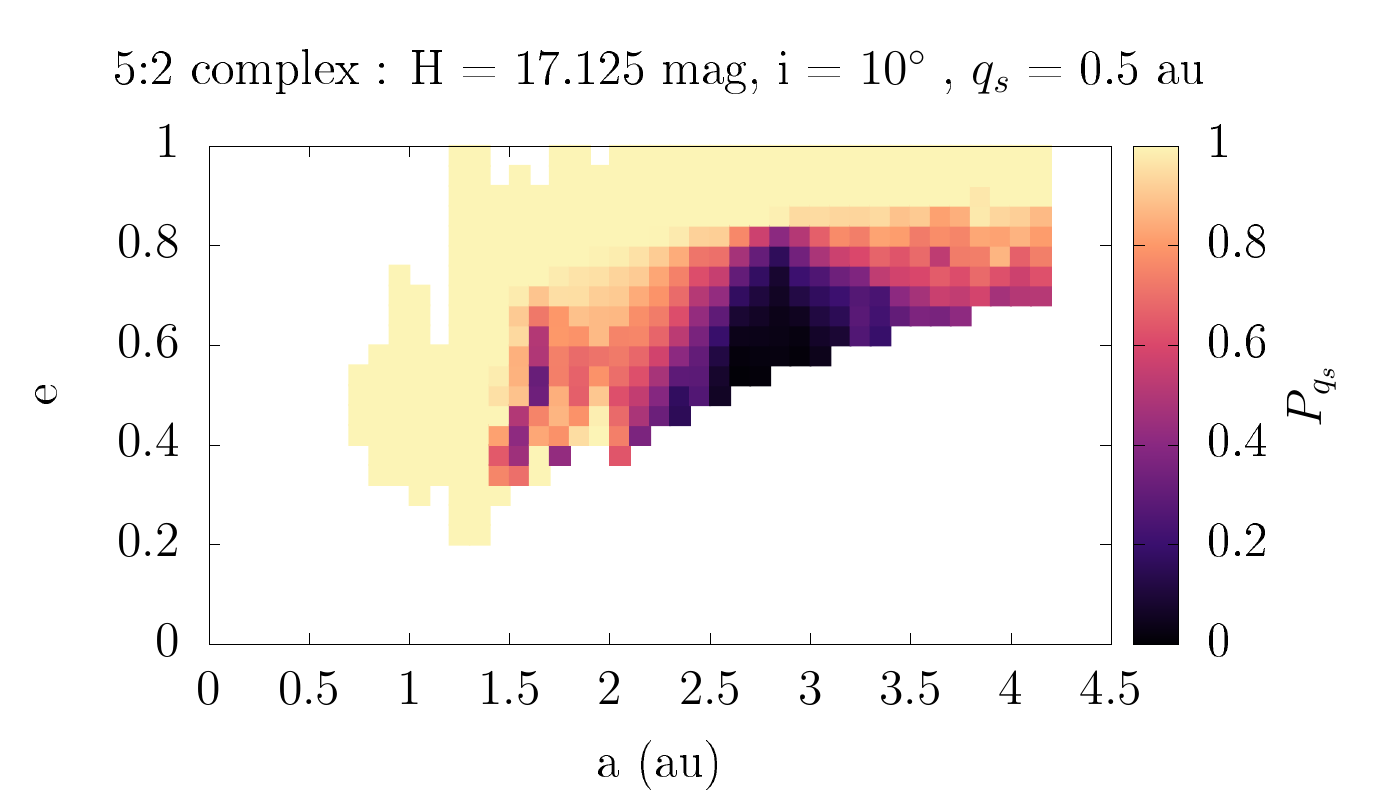}
\includegraphics[width=0.49\textwidth]{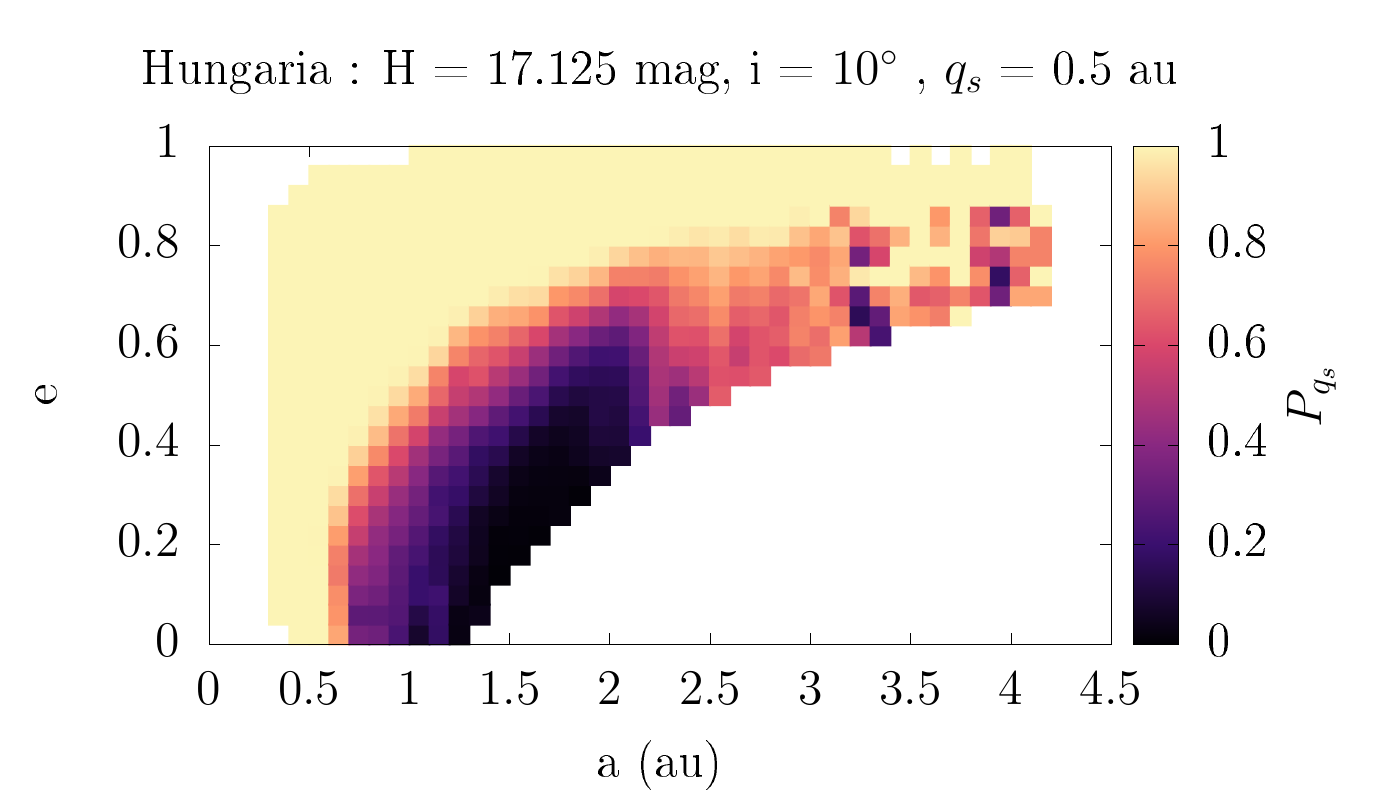}

\includegraphics[width=0.49\textwidth]{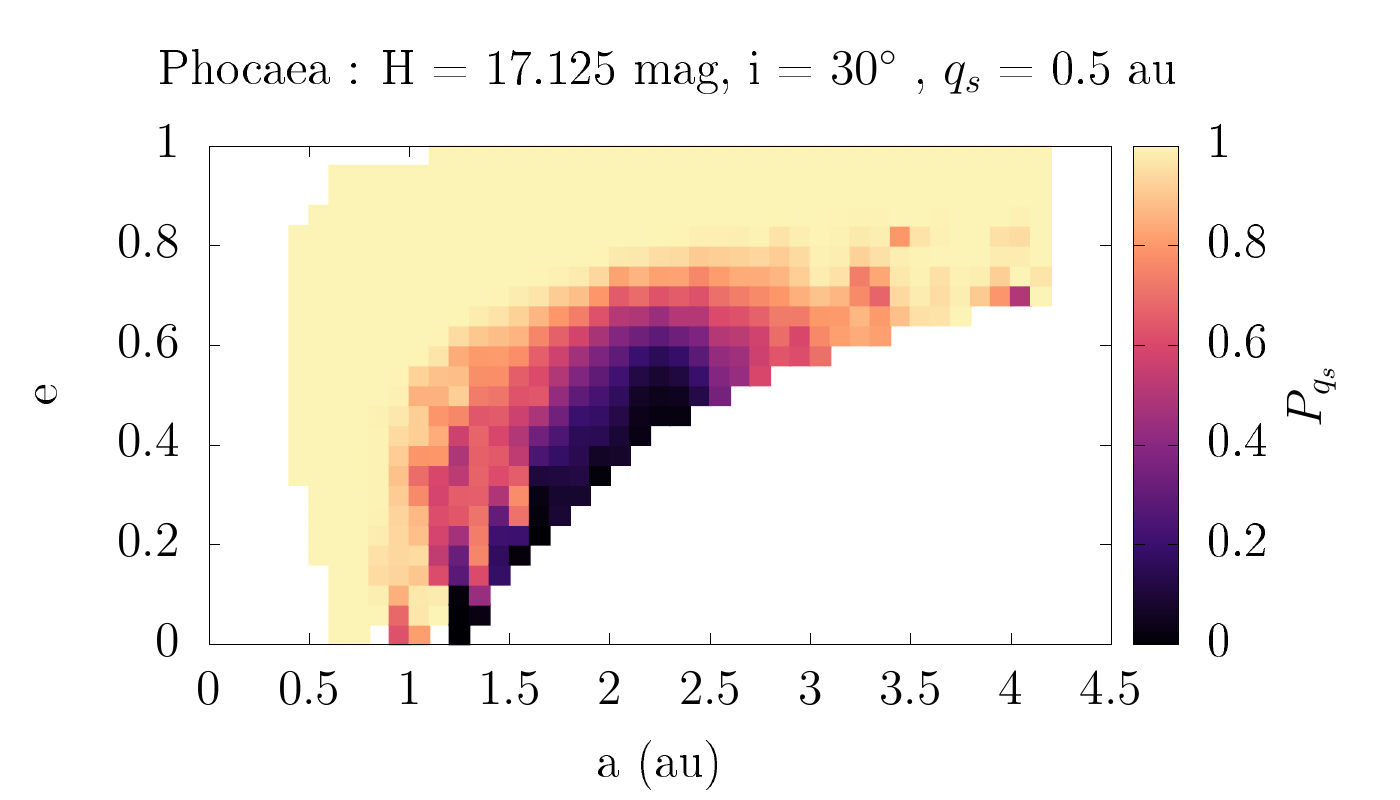}
\includegraphics[width=0.49\textwidth]{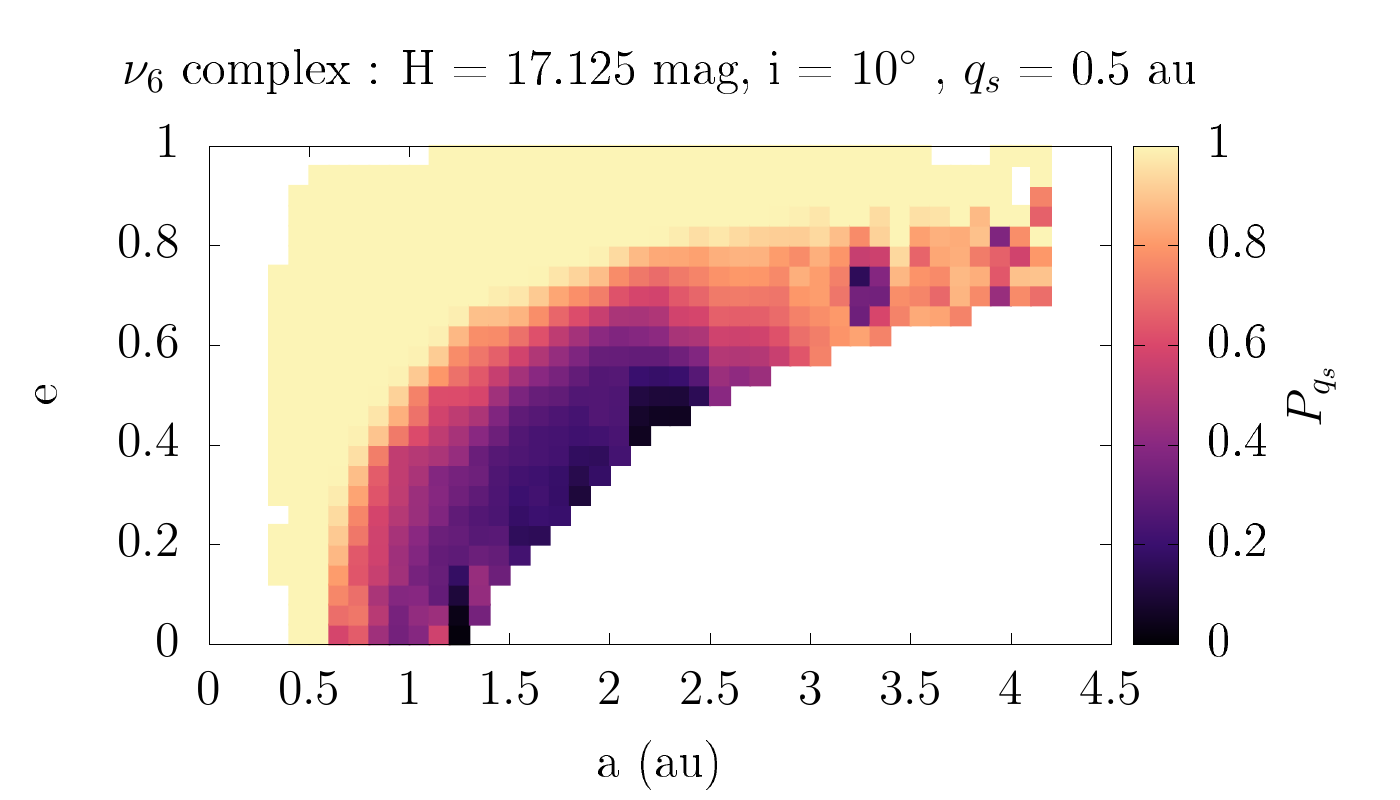}
\caption{The ER-specific distribution in the ($a,e$) plane that an asteroid coming (from top to bottom, left to right) from the 2:1, 3:1, 5:2 MMR complexes, the Hungaria and Phocaea groups, and the $\nu_6$ secular resonance complex has had in the past $q<0.5\au$. We use $H=17.125\hmag$ and $i=10\deg$ for all ERs, except for the Phocaeas for which we use $i=30\deg$. $P_{q_s}$ is lower in the location of the complexes. }
\label{fig:prob_sources}
\end{figure*}

Let us now take one step back and focus on the ER-specific probabilities, one for each of the six asteroidal ERs $p_{\text{ER}_{q_s}}(a,e,i)$. In Fig.~\ref{fig:prob_sources} we show the distribution in the ($a,e$) plane. The probability is lowest around the location of each four resonance complexes and one of the two asteroid families, Hungaria. The explanation is that this is their point of entrance from the MAB into the near-Earth region, and hence they have no prior history there. After entering, their orbits evolve and they leave these cells and it is not very likely that they return there later on.

Note that while this is true for a slice in $i$ equal to $10\deg$ for five out of the six ERs, it is not the case for the Phocaeas. As shown in \citet{Granvik2018},  NEAs originating in the Phocaea region enter the near-Earth region at larger $i$ and as a result this feature is apparent only if we select a higher inclination such as $i>20\deg$.

\section{Discussion}

\subsection{Fraction of near-Earth asteroids with small perihelion distances}

The most common sink for NEAs is disruption due to the proximity to the Sun. \citet{Marchi2009} noted that $\sim70$ per cent of NEOs end their lifetimes by falling on to the Sun. We have recorded the number of test asteroids that eventually come very close to the Sun and, more specifically, we have calculated which percentage of the $\sim70 000$ test asteroids, originating from all six ERs, have reached below $0.4$, $0.35$, $0.3$, $0.25$, $0.2$, $0.15$, $0.1$ and $0.05\au$ (that is, the first eight $q_s$ values) (Table~\ref{tab:percentages}).  

Apart from the contribution of each ER $\beta_{\text{ER}}(a,e,i,H)$, \citet{Granvik2018} provides us with the relative fraction of NEAs from each ER for asteroids with $17<H<25\hmag$, $\bar{\beta}(17<H<25)$. Thus, the linear combination between $\bar{\beta}(17<H<25)$ and the ER-specific percentages gives us the fraction of the total NEA population that, eventually, reaches below those $q_s$ values. As can be seen in the table, $\sim80$ per cent of the asteroids in the NEO population reach below $0.05\au$ during their evolution. However, in this calculation, we do not take into account the super-catastrophic disruption of NEAs at small $q$ due to irradiation from the Sun.

\begin{table*}
\centering
\caption{The ER-specific percentages of asteroids that reach below each of the first six $q_s$ values. The percentages corresponding to the total NEA population are weighted according to the contribution of each ER, averaged in the range $17<H<25\hmag$ \citep{Granvik2018}.}
\label{tab:percentages}
\begin{tabular}{lcccccccr}
\hline
Source & $q_s=0.05\au$ & $q_s=0.1\au$ &$q_s=0.15\au$ & $q_s=0.2\au$ & $q_s=025\au$ & $q_s=0.3\au$ & $q_s=0.35\au$ & $q_s=0.4\au$ \\
Region &(\%) &(\%) &(\%) &(\%) &(\%) &(\%) &(\%) &(\%) \\
\hline
2:1 complex     & 26.25 & 30.21 & 33.61 & 36.71 & 39.64 & 42.10 & 44.51 & 47.21 \\
3:1 complex     & 76.53 & 78.62 & 80.82 & 82.67 & 84.21 & 85.77 & 87.05 & 88.39 \\
5:2 complex     & 23.84 & 25.43 & 27.04 & 28.65 & 30.63 & 32.86 & 35.43 & 38.46 \\ 
Hungaria        & 80.61	& 82.80	& 84.60 & 86.31	& 87.81	& 89.33	& 90.53 & 91.74 \\
Phocaea        & 92.44	& 94.02 & 95.19 & 96.17 & 96.85 & 97.38	& 97.94 & 98.42 \\
$\nu_6$ complex & 82.87	& 84.70 & 86.76 & 88.46	& 90.05 & 91.36 & 92.68 & 93.86 \\
\hline
NEA population  & 79.11 & 81.12 & 83.19 & 84.95 & 86.51 & 87.98 & 89.27 & 90.55 \\

\hline
\end{tabular}
\end{table*}

Almost $80$ per cent of main-belt asteroids come into the near-Earth region though the $\nu_6$ and 3:1 resonances \citep{Granvik2018}. Typically, they are bright objects with $\sim88$ and $\sim86$ per cent, respectively, having a geometric albedo $p_V>0.1$ \citep{Morby2020}. In addition, these asteroids reach very small heliocentric distances with very large efficiency ($\sim80$ per cent). After taking into account their weighted contributions from each ER and their albedos, we find that of all the objects that eventually reach below $0.05\au$,  $\sim89$ per cent are bright, i.e., most likely S-type asteroids.

\begin{figure}
\centering
\includegraphics[width=0.48\textwidth]{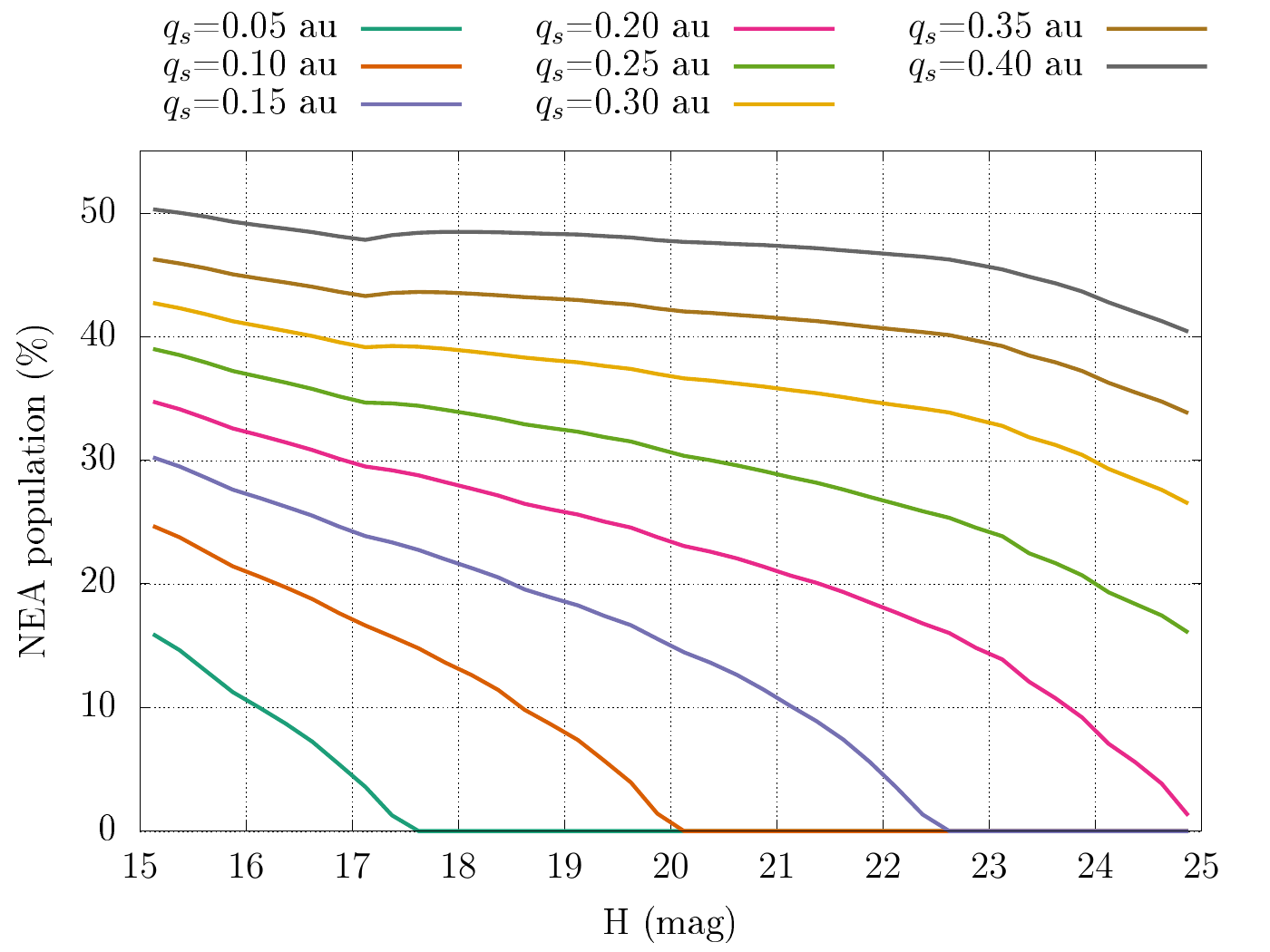}
\caption{The fraction of the NEA population as a function of $H$ that has had in its past $q$ smaller than the eight first $q_s$ values. }
\label{fig:population}
\end{figure}

In Fig.~\ref{fig:population} we plot the fraction of the steady state NEA population that, at some point in their past have had $q$ smaller than the first eight $q_s$ values. To produce these fractions, we binned the test asteroids by $H$, and, over all the non-empty ($a,e,i$) cells, we summed the products of $P_{q_s}\left(a,e,i,H\right)$ multiplied with the expected number of NEAs \citep{Granvik2018} and divided with the total number of expected objects in each particular $H$ bin. We find that for the same $q_s$, a smaller fraction of the population of small asteroids have reached below that threshold value during their evolution compared to larger ones. This is a consequence of our assumption of a super-catastrophic disruption of asteroids when their $q<\bar{q}_*$, which effectively prevents further orbital evolution. The $H$ frequency distribution of NEOs predicts a few orders of magnitude more objects with large $H$ values than with small so we still expect to find more small asteroids than large ones at  $q>0.2\au$. The small knee at $H\sim17\hmag$ is an artefact stemming from the fact that we consider the same source specific contribution to the total NEA population in the $15<H<17\hmag$ range, although there is a different number of expected asteroids for these $H$ bins. Note that this result cannot be directly compared to the results shown in Table~\ref{tab:percentages}; the analysis presented in Fig.~\ref{fig:population} shows how probable it is for a member of the NEA population, at any moment, to have had a $q$ that has reached below some $q_s$ value, while for Table~\ref{tab:percentages}, we consider the probability that any single test asteroid has reached below certain $q_s$ values by the end of its evolution and, also, do not take into account destruction of asteroids by thermal processes.

\subsection{Application to meteorite dropping fireballs}

Let us then turn to a practical example of how to use the results presented above. We took the orbital elements and their uncertainties of 25 meteorite dropping fireballs calculated by \citet{GranvikBrown2018}. For each meteorite fall, we located all the cells covered by the nominal ($a,e,i$) values and their uncertainties. Then we computed the average $P_{q_s}$ and $\tilde{T}_{q_b}$ coming from all those cells. In addition, we recorded their maximum and minimum values.

\begin{figure*}
\centering
\includegraphics[width=\linewidth]{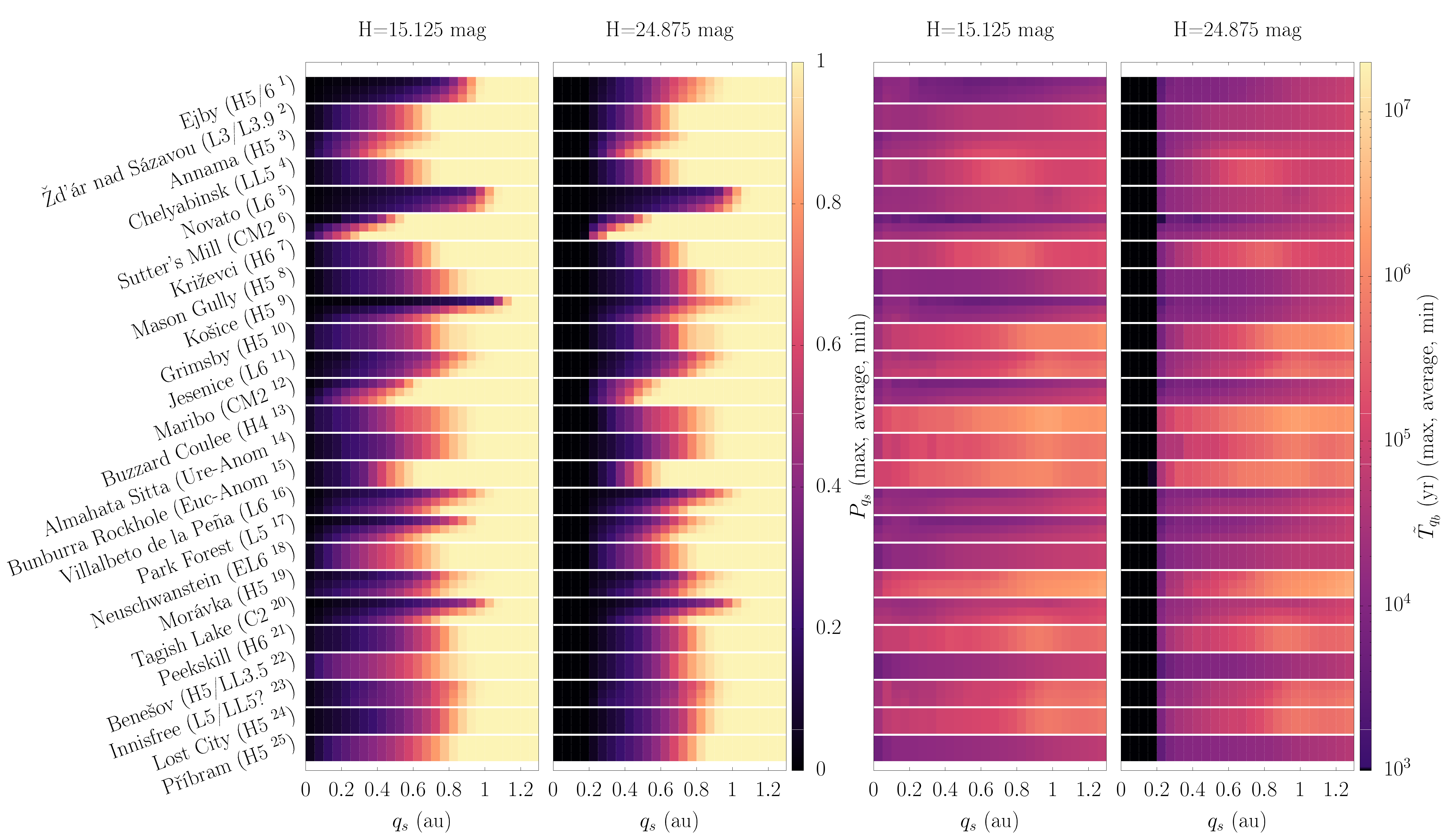}
\caption{Maximum, average, and minimum (left panels) $P_{q_s}$ and (right panels) $\tilde{T}_{q_b}$ of the immediate parent bodies to 25 meteorite falls. The results correspond to the two extremes of asteroid sizes in our $H$ range -- $H=15.125\hmag$ and $H=24.875\hmag$. Meteorite classifications taken from: $^{1}$\citet{Spurny2016a}, $^{2}$\citet{Spurny2015}, $^{3}$\citet{Kohout2016}, $^{4}$\citet{Kohout2014a}, $^{5}$\citet{Jenniskens2014}, $^{6}$\citet{Zolensky2014}, $^{7}$\citet{Lyon2014}, $^{8}$\citet{Dyl2016}, $^{9}$\citet{Ozdin2015}, $^{10}$\citet{McCausland2010a}, $^{11}$\citet{Bischoff2011}, $^{12}$\citet{Haak2012}, $^{13}$\citet{Hutson2009}, $^{14}$\citet{jen2009a}, $^{15}$\citet{Bland2009}, $^{16}$\citet{Llorca2005}, $^{17}$\citet{Simon2004}, $^{18}$\citet{Zipfel2010}, $^{19}$\citet{Borovicka2003b}, $^{20}$\citet{Brown2001a}, $^{21}$\citet{Wlotzka1993}, $^{22}$\citet{Spurny2014}, $^{23}$\citet{Kallemeyn1989}, $^{24}$\citet{Nava1971}, $^{25}$\citet{Tucek1961} .}
\label{fig:meteorites}
\end{figure*}

In Fig.~\ref{fig:meteorites}, we show $P_{q_s}$ and $\tilde{T}_{q_b}$ for the largest and the smallest $H$-bin centers in our grid, $H=15.125\hmag$ and $H=24.875\hmag$. The selected $H$ bins correspond to the two extreme cases available from our model for the size of the parent body. We note that meteorite falls, whose orbit determination yielded small uncertainties, i.e., that only cover one cell--the one corresponding to the nominal ($a,e,i$) values--have the same average, maximum, and minimum probabilities, and dwell times. 

\begin{figure}
\centering
\includegraphics[width=0.48\textwidth]{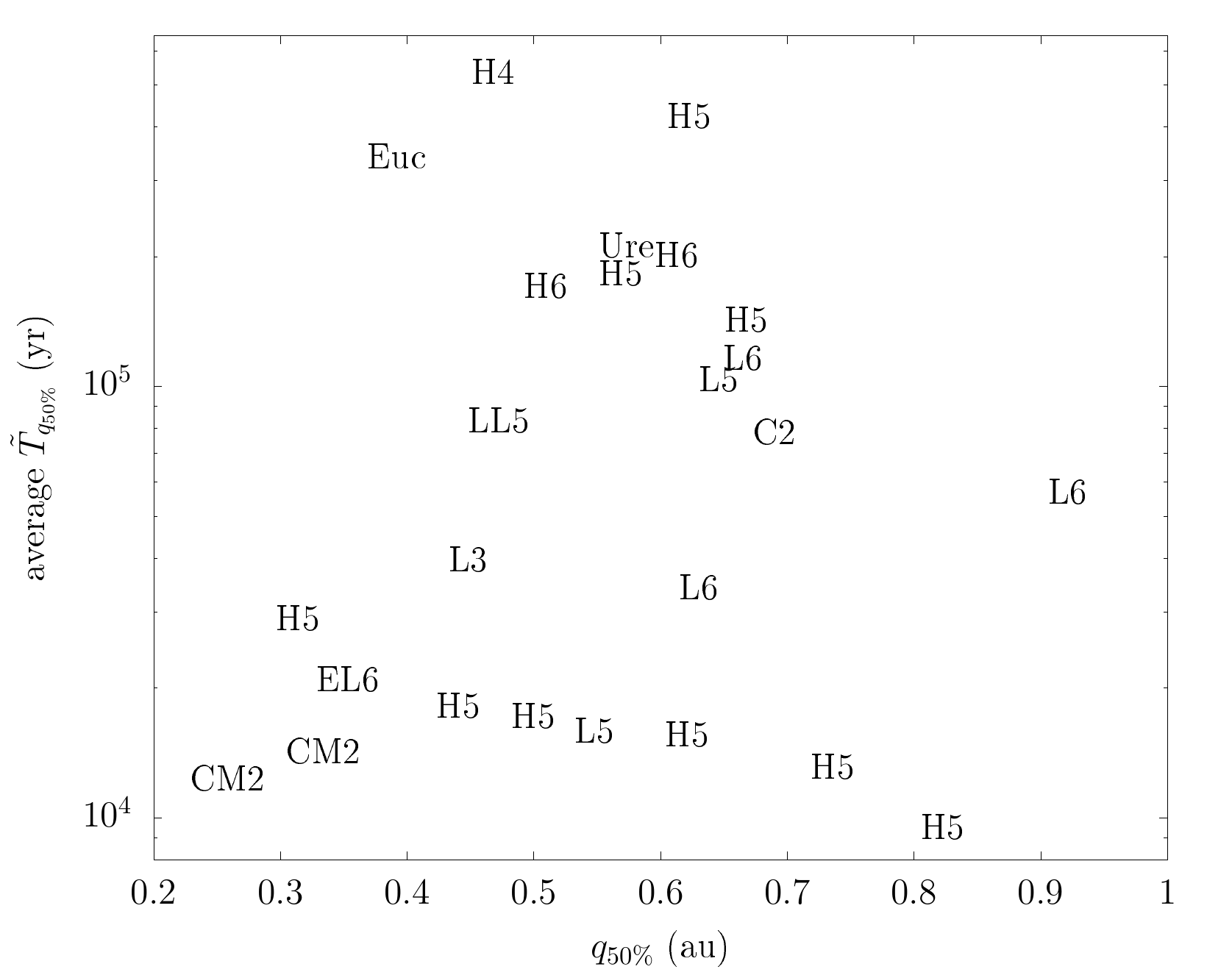}
\caption{The minimum $q$ corresponding to a probability of $50$ per cent versus the corresponding dwell time for 25 meteorite falls plotted with their classification as summarized by \citet{GranvikBrown2018}. The dwell time refers to the average values calculated from all the cells that are covered by the nominal values of $a,e,i$ and their uncertainties.}
\label{fig:med_meteorites}
\end{figure}

An overview of the probable past evolution of the immediate parent body of a meteorite fall can be seen by plotting the $q$ value which has a $50$ per cent probability to have been the minimum value that the parent body has reached ($q_{50\%}$), vs. the corresponding median dwell time at that $q$ ($\tilde{T}_{q_{50\%}})$. In Fig.~\ref{fig:med_meteorites}, we plot the distribution of each one of the 25 meteorites, according to their classification, in the ($q_{50\%}$ - $\tilde{T}_{q_{50\%}}$) plane. $q_{50\%}$ was calculated using the average $P_{q_s}$ mentioned above. We used the values that correspond to $H=15.125\hmag$ so this is the lower limit in terms of $q$ and upper limit in terms of dwell time. 

For ordinary chondrites there is no obvious correlation between, on one hand, the typical minimum $q$ and dwell time, and, on the other hand, meteorite classification (Fig.~\ref{fig:med_meteorites}). However, we note that there appears to be a dearth of meteorite falls with average $\tilde{T}_{q_{50\%}}<10^5\yr$, and $q_{50\%}\lesssim0.5\au$ or $q_{50\%}\gtrsim0.7\au$. Whereas the dearth of meteorite falls with $q_{50\%}\lesssim0.5\au$ and long dwell times is in agreement with the super-catastrophic destruction scenario \citep{Granvik2016,2019ApJ...873..104Y,2020AJ....159..143W}, we need a larger sample to be able to draw meaningful conclusions about the apparent dearth at $q_{50\%}\gtrsim0.7\au$.

One would expect to find the carbonaceous chondrites in the lower right quadrant of Fig.~\ref{fig:med_meteorites}, because dark asteroids have been shown to be more susceptible to super-catastrophic disruptions \citep{Granvik2016}. Whereas the model suggests that the parent bodies of the three carbonaceous chondrites likely had short dwell times--understandable assuming that a long exposure to solar irradiation at small $q$ would have destroyed them \citep{Granvik2016}--it is somewhat surprising to find the CMs in the lower \emph{left} quadrant, which indicates a small minimum $q$.

The small minimum $q$ for the CM chondrites suggests that their immediate parent bodies would have had to be at least a few meters in diameter to survive the minimum $q$ predicted or, if smaller, that their immediate parent bodies were fragments of larger objects that underwent super-catastrophic disruptions. The diameter of the immediate parent body of the Maribo meteorite has been estimated to be significantly smaller than one meter \citep{2012M&PS...47...30H}, suggesting that it could be a fragment from a super-catastrophic disruption event when it reached minimum $q$ or from a later disruption event. The diameter of the immediate parent body of the Sutter's Mill meteorite was estimated to be 1.8--3.5~m \citep{2012Sci...338.1583J}, which would indicate that it could have survived its nominal minimum $q$ of about $0.25\au$. However, laboratory studies have revealed that it has been exposed to temperatures exceeding 500C \citep{2014M&PS...49.1997Z}, which requires $q\sim0.15\au$. Such a small $q$ is still in line with our model (Fig.~\ref{fig:meteorites}) but suggests that the immediate parent body of the Sutter's Mill meteorite is, also, a fragment from a super-catastrophic disruption event when it reached minimum $q$ or from a later disruption event.

\section{Conclusions}

We have constructed a model in the form of a look-up table\footnote{\url{http://www.iki.fi/mgranvik/data/Toliou+_2021_MNRAS}} that enables us to assess for any asteroid with known orbital elements $a,e,i$ and $H$
\begin{itemize}
    \item the probability that it has in the past had $q$ smaller than some threshold value in the range $0<q_s\leq1.3\au$, and
    \item the amount of time it has spent having $q$ in a range defined by the threshold values $q_s$.
\end{itemize} 

The past evolution of $q$ of an asteroid can in turn be used to estimate its past thermal history. Thermal information can be of interest for studies of objects that can exhibit signs of extreme heating by irradiation from the Sun or of the potential parent bodies of meteorites whose composition and physical properties are under examination.  

We have identified similar trends as \citet{Marchi2009}, but our model supercedes their model in that it is based on a new and updated NEO population model. In addition, our model not only concerns the currently known NEAs, but cover the entire near-Earth region, which allows applying it to any NEO, also those that will be discovered in the future. Finally, our model also accounts for the super-catastrophic disruption of NEAs close to the Sun, and has an explicit dependence on $H$.

We apply our model to 25 meteorite falls and find that carbonaceous chondrites typically have short dwell times at small $q$ whereas the dwell times of ordinary chondrites range from 10 thousand years to half a million years. A dearth of meteorite falls with long dwell times and small minimum $q$ is in agreement with a super-catastrophic disruption of asteroids at small $q$.

\section*{Acknowledgements}

We thank the anonymous reviewer for the rapid turnaround as well as the constructive criticism that improved the paper. AT, MG, and GT acknowledge funding from the Knut and Alice Wallenberg Foundation, and MG also from the Academy of Finland.

\section*{Data Availability}

No new data were generated in support of this research.



\bibliographystyle{mnras}
\bibliography{bibliography} 








\bsp	
\label{lastpage}
\end{document}